\documentclass[twocolumn,prd,aps,superscriptaddress]{revtex4-1}
\usepackage{graphicx}
\usepackage{bm}
\usepackage{color}
\usepackage{amsmath}
\usepackage{amsfonts}
\usepackage{verbatim}
\usepackage{amssymb}
\usepackage{graphicx}
\usepackage{epstopdf}
\usepackage{mathrsfs}
\usepackage{epsfig}
\usepackage{slashed}
\usepackage{bbold}

\providecommand{\U}[1]{\protect\rule{.1in}{.1in}}
\begin{document}

\title{
Thermal Casimir effect for a Dirac field on flat space with a nontrivial circular boundary condition
}

\author{Jo\'{a}s Ven\^{a}ncio}
\email[]{joasvenancioufpe@gmail.com}
\affiliation{Departamento de F\'{\i}sica, Universidade Federal de Pernambuco, Recife, Pernambuco  50740-560, Brazil.}
\author{Lameque Filho}
\email[]{lmsf@academico.ufpb.br}
\affiliation{Departamento de F\'{i}sica, Universidade Federal da Para\'{i}ba, Jo\~{a}o Pessoa, Caixa Postal 5008, Brazil.}
\author{Herondy Mota}
\email[]{hmota@fisica.ufpb.br}
\affiliation{Departamento de F\'{i}sica, Universidade Federal da Para\'{i}ba, Jo\~{a}o Pessoa, Caixa Postal 5008, Brazil.}
\author{Azadeh Mohammadi}
\email[]{azadeh.mohammadi@ufpe.br}
\affiliation{Departamento de F\'{\i}sica, Universidade Federal de Pernambuco, Recife, Pernambuco  50740-560, Brazil.}

\begin{abstract}

This work investigates the thermal Casimir effect associated with a massive spinor field defined on a four-dimensional flat space with a circularly compactified spatial dimension whose periodicity is oriented along a vector in $xy$-plane. 
We employ the generalized zeta function method to establish a finite definition for the vacuum free energy density. This definition conveniently separates into the zero-temperature Casimir energy density and additional terms accounting for temperature corrections.
The structure of existing divergences is analyzed from the asymptotic behavior of the spinor heat kernel function and removed in the renormalization by subtracting scheme. The only non-null heat coefficient is the one associated with the Euclidean divergence. 
We also address the need for a finite renormalization to treat the ambiguity in the zeta function regularization prescription \text{associated} with this Euclidean heat kernel coefficient and ensure that the renormalization procedure is unique. The high- and low-temperature asymptotic limits are also explored. 
In particular, we explicitly show that free energy density lacks a classical limit at high temperatures, and the entropy density agrees with the Nernst heat theorem at low temperatures.
\end{abstract}

\maketitle

\vspace{3cm}
\section{Introduction}

The Casimir effect is a fascinating quantum phenomenon initially proposed by H. Casimir in 1948 \cite{Casimir}. 
In its standard form, such an effect establishes that 
two parallel, electrically neutral conducting plates in close proximity experience an attractive force inversely proportional to the fourth power of the distance between them. This attraction arises from alterations in vacuum fluctuations of the electromagnetic field.
Since this force between the plates is extremely weak, the Casimir effect was initially perceived as a theoretical curiosity. M. Sparnaay conducted the pioneering experimental attempt, however with low precision,  to detect this effect in 1958 \cite{Sparnaay}. It was only confirmed decades after by several high-accuracy experiments \cite{Bressi2002, Lamoreaux, Lamoreauxx, Mohideen}. Since then, spurred by the progress 
in theories of particles and fields, the Casimir effect has been investigated in increasingly complicated configurations, not only due to its theoretical and mathematical aspects but also due to the countless technological applications arising from the macroscopic manifestation of a fully quantum effect \cite{Bordag, Klimchitskaya, Mostepanenko, Ford1975, Dowker 1976, Dowker1978, Appelquist1983, Hosotani1983, Brevik2002, Zhang2015, Henke2018, Bradonjic2009, Peng2018, Pawlowski2013, Gambassi2009, Machta2012, Milton2019}. 
A thorough review concerning the Casimir effect is presented in Refs. \cite{Klimchitskaya2009, Milton2001}.

Although originally associated with the electromagnetic field, the Casimir effect is not an exclusive feature of this particular field. Other fields, for instance, scalar and spinor fields, and gauge fields (Abelian and non-Abelian), can exhibit analogous phenomena under nontrivial boundary conditions
\cite{Stokes2015, Farina2006, Muniz2018, Cunha2016, Mobassem2014, Bytsenko2005, Pereira2017, Photon2001, Chernodub2018, Edery2007}. However, among the vast literature concerning the Casimir effect, the majority of the investigations have been focused on scalar fields. 
The reason for this is not conceptual but, most likely, the more significant technical complexity involved in the formalism needed to treat spinor fields, for instance.

Spinor fields play an important role in many branches of physics since they represent fermion fields. Additionally, they carry the fundamental representation of the orthogonal group, making spinors the building block of all other representations of this group. In this sense, spinors are the most fundamental entities of a space endowed with a metric \cite{Cartan1966, Benn1987, JoasBook2019}. In particular, studying vacuum energy associated with the quantized version of these fields sets a scenario for which the physics involved is quite rich.

The presence of divergencies is an inherent feature of vacuum energy when calculated with the quantum field theory (QFT) techniques. Knowing how to deal with them is challenging in general. This special concern has resulted in the development of regularization and renormalization techniques in mathematical physics, which can be applied to remove the divergences associated with the calculations involved in the Casimir effect \cite{Oikonomou2010,Cheng2010,Cavalcanti2004,Elizalde2008}. 
This study concentrates explicitly on a robust and elegant regularization method employing the generalized zeta function. This function is constructed from the eigenvalues of a differential operator, which governs the quantum field dynamics \cite{Elizalde1995, Hawking1977, Elizalde2012}.  
The divergencies are typically introduced in the partition function in QFT by the determinant of the operator, which is an infinite product over all eigenvalues, and encoded into the generalized zeta function
\cite{Elizalde1994}. 
Once we obtain the partition function, the canonical ensemble establishes the formal connection with thermodynamics. It facilitates the calculation of free energy, which allows for considering temperature corrections to the vacuum energy. \cite{Basil1978, Plunien1986, Kulikov1988, Maluf2020}. The structure of the existing divergences in these calculations typically involves examining the asymptotic behavior of the two-point heat kernel function associated with the relevant operator, as considered in M. Kac's seminal paper \cite{Kac1966} and further explored in \cite{Elizalde1994, Bordag2000, Vassilevich2003, Kirstein2010}. 
This zeta function investigation predominantly focuses on Laplace-type operators associated with scalar fields, with comparatively less emphasis on Dirac-type operators associated with spinor fields \cite{Branson1992A, Branson1992B}.

One potential explanation for this disparity is the requirement for the considered operator, which governs the propagation of the quantum field under specified boundary conditions, to be self-adjoint. The self-adjointness is necessary for the construction of zeta and heat kernel functions. 
The most common boundary conditions, widely used in the Casimir effect for the scalar field, are Dirichlet and Neumann ones. 
However, these conditions do not directly extend to spinor fields due to the first-order nature of Dirac operators. Instead, the bag model boundary conditions first presented in Refs. \cite{Chodos1974, Johnson1975} make the Dirac operator formally self-adjoint. This was also investigated in Ref. \cite{Arrizabalaga2017} recently. In particular, the Casimir effect for spinor fields under bag model boundary conditions has been addressed in Ref. \cite{Mamayev1980} and for Majorana spinor fields with temperature corrections in Refs. \cite{Oikonomou2010, Cheng2010, Erdas2011,Elizalde2012Maj}. Alternative methods to maintain the self-adjoint nature of the Dirac operator have also been explored. For example, the Casimir effect involving spinor fields confined by a spherical boundary has been examined in Refs. \cite{Bender1976, Elizalde1998} using the zeta function method. This approach was recently extended to include a spherically symmetric $\delta$-function potential in \text{Ref. \cite{Fucci2023}}.  Furthermore, Elko fields, which are spinor fields satisfying a Klein-Gordon-like equation, allow for the imposition of boundary conditions similar to those used for scalar fields. The finite temperature Casimir effect for Elko spinor fields in a field theory at a Lifshitz fixed point is discussed in Refs. \cite{Pereira2017, Pereira2019, Maluf2020}.

Boundary conditions play a pivotal role in the exploration of the Casimir effect. Interestingly, it is possible to induce boundary conditions through identification conditions in spaces with nontrivial topology, thereby eliminating the need for material boundaries.
Such topologies induce boundary conditions on the quantum fields that distort the corresponding vacuum fluctuations, such as a material boundary does, producing a Casimir-like effect \cite{Klimchitskaya, Milton2001}. 
The Casimir effect for different types of fields and boundary conditions in spaces with nontrivial topology has been addressed in Refs. \cite{Mostepanenko2011, HerondyJunior2015, Mohammadi2022, Herondy2023, Farias2020, Xin-zhou, Zhai, Li, Xin2011}.  

In the present work, we have delved into the thermal Casimir effect using the generalized zeta function approach for a massive spinor field defined on a four-dimensional flat space with a circularly compactified spatial dimension, whose periodicity is oriented not along a coordinate axis as usual, but along a vector in the ${xy\text{-plane}}$, dubbed compact vector.
This space introduces a topological constraint that imposes a spatial anti-periodic boundary condition along the compact vector on the spinor field. Up to a coordinate origin redefinition, this condition is referred to as the anti-helix condition in Ref. \cite{Xiang2011}, where the authors investigated the zero-temperature Casimir effect for spinor fields induced by the helix topology. However, to our knowledge, a study that adds thermal effects induced by this topology in the spinor field context has not appeared in the literature. The calculations conducted in this study not only extend the findings from Ref. \cite{Xiang2011} to finite temperature but also revisit the results from Ref. \cite{Bellucci2009} in a limiting case. Additionally, our study serves as a spinor extension of the thermal Casimir effect studied in Ref. \cite{Giulia2021}, which focused on scalar fields subjected to a helix boundary condition.

The structure of this paper is organized as follows. Section \ref{Path integral} provides a general expression for the partition function associated with a massive Dirac field defined on a space endowed with a flat Euclidean metric, in the path integral representation. In Section \ref{Zeta function of an operator}, we outline the mathematical framework employed to compute the vacuum-free energy using the generalized zeta function method. 
This method involves imposing an anti-periodic condition on the Dirac field in imaginary time $t$ and analyzing existing divergences based on the asymptotic behavior of the spinor heat kernel. 
In particular, we discuss the presence of ambiguities in the zeta function regularization due to nonzero heat kernel coefficients and the necessity of requiring vacuum energy to renormalize to zero for large masses.
In Section \ref{Helix topology}, we derive the spinor heat kernel two-point function and the Casimir energy density, incorporating temperature corrections induced by the anti-periodic boundary condition along the compact vector. We also analyze the low- and high-temperature asymptotic limits.  Finally, Section \ref{Conclusion} provides a summary of the paper, highlighting the distinctions between the spinor and scalar cases. Throughout this paper, we adopt the natural units where $c = \hbar = k_B = 1$.

\vspace{-0.09mm}

\section{Path integrals}\label{Path integral}

To illustrate the use of the generalized zeta function method in quantum field theory (QFT), we revisit some known underlying facts. 
In the path integral formulation, the one-loop partition function associated with a complex matter field $\Psi$ (and its conjugate $\bar{\Psi}$) can be obtained from the following source-free generating functional
\begin{equation}\label{path integral}
Z = \int  \mathcal{D}\bar{\Psi} \mathcal{D}\Psi \, e^{i\mathcal{S}(\Psi, \bar{\Psi})} ,
\end{equation}
where $\mathcal{D}\bar{\Psi} \mathcal{D}\Psi$ stands for the integration measure over the field space, whose dynamics is described by the action ${\mathcal{S}}(\Psi, \bar{\Psi})$. Such representation provides a straightforward method for introducing temperature into QFT. This can be achieved by defining a Euclidean action $\mathcal{S}_E(\Psi, \bar{\Psi})$ through a rotation in the complex plane, known as Wick rotation,
with the fields satisfying periodic (for scalar fields) or anti-periodic (for spinor fields) conditions in imaginary time with period $\beta$. 
In this Euclidean approach to QFT, $Z$ is the one-loop partition function for a canonical ensemble at the temperature $T = \beta^{-1}$.

\subsection{Spinor fields}

We can start with the path integral for spinor fields. 
Let $\{\boldsymbol{e}_{a}\}~(a = 1, 2, 3, \ldots, N)$ be an orthonormal frame field that spans  $M= (\mathbb{R}^{N}, \boldsymbol{g})$, a $N$-dimensional space endowed with a flat Euclidean metric $\boldsymbol{g}$ whose components with respect to basis $\{\boldsymbol{e}_{a}\}$ are
\begin{equation}\label{metric component}
    \boldsymbol{g}(\boldsymbol{e}_{a}, \boldsymbol{e}_{b}) =  \delta_{ab} ,
\end{equation}
where $\delta_{ab}$ is the Kronecker delta.
That is, the space can be covered by cartesian coordinates ${\{x^{a}\} = \{t, x,  y, \dots, z\}}$ such that the line element on $M$ is given by 
\begin{align}
   ds^2 = dt^2 + dx^2 + dy^2 + \cdots + dz^2 .
\end{align}
The imaginary time coordinate $t$ is compactified into a finite length equal to the inverse of temperature $\beta$, so that $M$ is closed in the $t$-direction. This is equivalent to consider spinor fields on 
$M = \mathbb{S}^1 \times \mathbb{R}^{N-1}$ satisfying anti-periodic boundary conditions. The associated action has the form
\begin{equation}\label{Euclidean action}
    \mathcal{S}_E(\Psi, \bar{\Psi}) =  \int_{0}^{\beta} dt\int d^{N-1} x\, \sqrt{g} \, \bar{\Psi}(x)\slashed{D}(m)\Psi(x) , 
\end{equation}
where $g$ is the metric determinant and $\slashed{D}(m)$ is the standard skew-adjoint Dirac operator, $\slashed{D} = \gamma^{a}\partial_{a}$, in the presence of a mass term
\begin{equation}
  \slashed{D}(m) = \gamma^{a}\partial_{a} + m , \quad a = 1, 2, 3, \ldots, N.
\end{equation}
 The frame $\{\boldsymbol{e}_a\}$ can be faithfully represented by the Dirac matrices $\{\gamma_a\}$ that generate the Clifford algebra over $M$
\begin{equation}
    \gamma_a \gamma_b + \gamma_b \gamma_a = 2 \, \boldsymbol{g}(\boldsymbol{e}_{a}, \boldsymbol{e}_{b}) \mathbb{1}. 
\end{equation}
In Euclidean signature, the Dirac matrices defined above are Hermitian, denoted by ${\gamma_{a}^{\dagger} = \gamma_{a}}$, and the conjugate spinor $\bar{\Psi}$ is simply the Hermitian conjugate of $\Psi$, written as ${\bar{\Psi} = \Psi^{\dagger}}$.
Since the dimension of the spinor space in $N$ dimensions is $[N/2]$ (the floor of the number $N/2$), $\{\gamma_a\}$ and $\mathbb{1}$ stand for $2^{[D/2]} \times 2^{[D/2]}$ matrices. In four dimensions ($N=4$), for instance, they are $4 \times 4 $ matrices. 
The spectral theory of general first-order differential operator of Dirac type can be found in Refs.  \cite{Branson1992A,Branson1992B}.
 

Our goal is to solve the integral \eqref{path integral}. To accomplish this, we can expand the spinor fields $\Psi$ and $\bar{\Psi}$ in terms of four-component complete orthonormal sets of Dirac spinors $\psi_j$:
\begin{align}
    \Psi(x) = \sum_j \Psi_j \psi_j(x), \label{Field expansion Psi}\\
     \bar{\Psi}(x) = \sum_j \bar{\Psi}_j \psi^{\dagger}_j(x),  \label{Field expansion Psi bar}
\end{align}
The coefficients $\Psi_j$ and $\bar{\Psi}_j$ are independent Grassmannian variables, and the index $j$ labels the field modes. 
The spinors $\psi_j$ are eigenfunctions of $\slashed{D}$ with eigenvalues determined by the equation 
\begin{equation}
    \slashed{D} \psi_j= i \lambda_{j} \psi_j, \quad \forall ~ \lambda_j \in \mathbb{R} ,
\end{equation}
and satisfy the following orthonormality and completeness relations 
\begin{align}
    \int d^{N}x\,\sqrt{g} \, \psi^{\dagger}_j(x) \psi_k(x) = \delta_{jk} ,\label{orthonormality}\\
    \sum_j \psi_j(x)  \psi^{\dagger}_j(x')  = \delta(x-x')\, \mathbb{1} \label{orthonormality matricial},
\end{align}
where $\delta(x-x')$ is Dirac delta-function in the Euclidean coordinates $\{x, x'\}$. 
Taking into account the orthonormality property \eqref{orthonormality} and the field expansions \eqref{Field expansion Psi} and \eqref{Field expansion Psi bar}, the action \eqref{Euclidean action} can be put into the diagonal form
\begin{align}
     \mathcal{S}_E(\Psi, \bar{\Psi}) &  = \sum_j \lambda_j(m) \bar{\Psi}_j {\Psi}_j ,
\end{align}
where $\lambda(m)$ is given by:
\begin{align}
   \lambda_j(m)  =  i \lambda_j + m . 
\end{align}

Now, under the decompositions \eqref{Field expansion Psi} and \eqref{Field expansion Psi bar}, the anti-periodic functional integral over the fields can be written in terms of $\Psi_j$ and $\bar{\Psi}_j$ as
\begin{align}
  \int_{\text{anti-periodic}}  \mathcal{D} \bar{\Psi}  \mathcal{D} \Psi= \int \prod_j \frac{1}{\mu} \,  d\bar{\Psi}_j d\Psi_j, 
\end{align}
in which an arbitrary scale parameter $\mu$ has been introduced. An interesting discussion on the meaning of $\mu$ can be consulted in \cite{Elizalde1990}.
By using the fact that the integration rules for Grassmannian degrees of freedom are
    \begin{equation}\label{Grassmann rules}
    \int d \Psi_j = 0 \quad \text{and} \quad \int \Psi_j d \Psi_j = 0 ,
\end{equation}
which must be equally satisfied by $\bar{\Psi}_j$, we are eventually led to the following result
\begin{align}\label{Granssmann integral}
   & {\,}\int d \bar{\Psi}_j d \Psi_j e^{-\lambda_{j}(m)\bar{\Psi}_j \Psi_j} \nonumber\\
   = &\int d \bar{\Psi}_j d \Psi_j \left[1 - \lambda_{j}(m)\right] \bar{\Psi}_j \Psi_j =  \lambda_{j}(m) .
\end{align}
The exponential series' quadratic and higher-order powers vanish identically due to Grassmannian anticommutative properties. 
Assuming that \text{Eqs. \eqref{Field expansion Psi}-\eqref{Granssmann integral}} hold, the path integral \eqref{path integral} over the Grassmann-valued Dirac spinors $\Psi$ and $\bar{\Psi}$ gives the one-loop functional determinant of the operator $\slashed{D}(m)$ with a positive exponent, as follows
\begin{align}\label{Functional determinant}
Z & = \int \prod_j \frac{1}{\mu}\,d \bar{\Psi}_j d \Psi_j\,e^{-\sum_j\lambda_{j}(m)\bar{\Psi}_j \Psi_j} \nonumber\\
  &= \prod_{j} \frac{\lambda_{j}(m)}{\mu}   = \text{det}\left[\frac{\slashed{D}(m)}{\mu} \right].
\end{align}
Note that the above functional determinant is divergent because of infinite product over the eigenvalues. This divergence indicates a need for some regularization procedure. In this paper, we will adopt a powerful and elegant regularization technique that utilizes the so-called generalized zeta function, the zeta function of an operator.

\section{Generalized zeta function}\label{Zeta function of an operator}


Let $L$ be a positive-definite self-adjoint second-order elliptic differential operator, i.e. the eigenvalues $\lambda_j$ of $L$ are real and non-negative. The zeta function associated with the operator $L$ is defined as
\begin{equation}\label{generalized zeta}
    \zeta_{L}(z) = \sum_j \lambda_{j}^{-z} , 
\end{equation}
where the sum over $j$ means the sum over the spectrum of $L$.
In $N$ dimensions, the serie \eqref{generalized zeta}  will converge for ${\text{Re}(z) > N/2}$  
and 
can be analytically continued for the other values of $z$ \cite{Seeley1967}. In particular, it is regular at $z = 0$.

Now, we can use the zeta function above to provide a regularized version of the ill-defined product of all eigenvalues. Taking the exponential of the derivative of the zeta function with respect to $z$, evaluated at $z=0$, the zeta-function regularized determinant can be defined by the relation 
\begin{equation}\label{regulated determinant}
    \text{ln~det}L = \sum_j \text{ln}\lambda_j := -\zeta_{L}'(0),
\end{equation}
where $\zeta_{L}'(z)$ stands for the derivative of $\zeta_{L}(z)$ with respect to $z$. The definition \eqref{regulated determinant} is well defined because the zeta function is regular at $z = 0$, and encodes all divergences present in the sum $\sum_j \text{ln}\lambda_j$. 

Defined previously as a series over the eigenvalues of an operator, the zeta function admits also an integral representation by making a Mellin transform, that is
\begin{align}\label{integral representation}
   \zeta_{L}(z)  = \frac{1}{\Gamma(z)} \int_{0}^{\infty} d\tau\,\tau^{z-1} K_{L}(\tau) ,
\end{align}
where $K_{L}(\tau)$ is a spectral function called global heat kernel, defined as
\begin{align}\label{heat kernel trace}
    K_{L}(\tau) = \text{Tr}\left(e^{-\tau L} \right) .
\end{align}
with Tr standing for the trace operation. In the case of the operator $\slashed{D}^2(m)$ which is a $2^{[N/2]} \times 2^{[N/2]}$ matrix in the spinor indices, Tr should be understood with an extra factor $2^{[N/2]}$ included. 
Besides that, being $\lambda_j$ the eigenvalues of the operator $L$, we can rewrite ${ \text{Eq.}~\eqref{heat kernel trace}}$ as 
\begin{align}
    K_{L}(\tau) = \sum_j e^{-\tau \lambda_j},
\end{align}
which diverges for $\tau \rightarrow 0$. In general, the structure of the divergences present in the zeta function can be accessed from the asymptotic behavior of the heat kernel for small $\tau$. For $\tau \rightarrow 0$, the heat kernel admits the following expansion
\begin{align}\label{heat equation expansion}
    K_{L}(\tau) \sim \frac{1}{(4 \pi \tau)^{N/2}}\sum_{p} c_{p}(L)  \,\tau^{p}, ~   p = 0, \frac{1}{2}, 1,\frac{3}{2}, \ldots ,
\end{align}
where $c_p := c_{p}(L)$ are the heat kernel coefficients.  
To review many of the basic properties of the heat kernel method in QFT, including some historical remarks, we refer to \cite{Vassilevich2003, Vassilevich2011, Gilkey1995}. 


It is now possible to obtain a link between the one-loop partition function and the generalized zeta function. Using the cyclic property of the trace and the fact that the hermitian chiral matrix $\gamma^{5}$ (denoted this way independently of the dimension) anticommutes with all Dirac matrices $\gamma^a$,  we can show the important property
\begin{align}
    \text{Tr ln} \left[\slashed{D}(m)\right] & =  \text{Tr ln}\left[ \gamma^{5}\slashed{D}(m)\gamma^{5}\right] =  \text{Tr ln} \left[\slashed{D}^{\dagger}(m)\right] \nonumber\\
    & = \frac{1}{2} \left\{ \text{Tr ln}\left[\slashed{D}(m)\right] +  \text{Tr ln}\left[\slashed{D}^{\dagger}(m)\right]\right\} \nonumber\\
    & =  \frac{1}{2} \text{Tr ln}\left[\slashed{D}^2(m)\right] , \label{trace property} 
    \end{align}
where $\slashed{D}^2(m)$ is the negative of the spinor Laplacian on $M$, $\slashed{D}^2 = \gamma^{a}\gamma^{b} \partial_{a}\partial_{b}$, in the presence of the mass
\begin{align}
    \slashed{D}^2(m) = -\slashed{D}^2 + m^2.  
\end{align}
Note in particular that the spinors $\psi_j$ are eigenfunctions of $\slashed{D}^{2}(m)$ with non-negative eigenvalues
\begin{align}\label{Eigenvalue equation fo D^2}
   \slashed{D}^{2}(m) \psi_j = \left( \lambda_{j}^{2} + m^2  \right)\psi_j .
\end{align}   
Employing the identity
\begin{equation}
    \text{det} L  = e^{\text{Tr~ln} L} ,
\end{equation}
one can derive from Eq. \eqref{trace property} the important relation
\begin{equation}\label{square root of the Dirac operator}
   \text{ln}\,\text{det} \left[\slashed{D}(m) \right] = \frac{1}{2}\, \text{ln}\,\text{det} \left[\slashed{D}^{2}(m)\right],
\end{equation}
establishing the massive extension of the relation between the determinant of the Dirac operator and the square root of the determinant of its associated Laplace-type operator. 
From Eqs. \eqref{generalized zeta}, \eqref{regulated determinant} and \eqref{square root of the Dirac operator}, 
the zeta-function regularization allows us to write the one-loop partition function as follows \cite{Bytsenko1992}
\begin{equation}\label{Log Z}
\text{ln}Z = -\frac{1}{2} \left[\zeta_{\slashed{D}^{2}(m)}'(0) + \text{ln}(\mu^2)  \zeta_{\slashed{D}^{2}(m)}(0)\right] ,
\end{equation}
which has the same structure as the scalar case, up to a global sign \cite{Hawking1977,Giulia2021}. This is expected since we are working with the zeta function associated with operator $\slashed{D}^{2}(m)$, which is of Laplace type, instead of $\slashed{D}(m)$. 

With the expression \eqref{Log Z}, one can obtain the free energy $F$, defined as \cite{Landau1980}
\begin{equation}\label{free energy}
   F = -\frac{1}{\beta}\, \text{ln} Z ,
\end{equation}
which is needed for the derivation of the Casimir energy at finite temperature. A  thermodynamics quantities closely related to the free energy is the entropy, defined as
\begin{align}\label{Casimir entropy}
    S & = - \frac{\partial F}{\partial T} , 
\end{align}
which, as we will see later, satisfies the third law of thermodynamics (the Nernst heat theorem).

Although the zeta-function method encodes all divergences present in the functional determinant, the structure of these divergences, however, plays a central role in the renormalization procedure. Let us now utilize this mathematical machinery to discuss a generic case of the Casimir energy associated with the spinor field in four dimensions. To achieve our purpose, it is convenient to decompose the time dependence of the spinor field in the Fourier basis, namely
\begin{equation}\label{time decomposition}
    \psi_j(x) = e^{-i\omega_{n} t} \chi_{\ell}(\boldsymbol{r}) ,
\end{equation}
stemming from the fact that $\partial_t$ is an obvious Killing vector field of our metric, where $\ell$ is a generic index denoting the spatial quantum modes of the field.  Imposing the anti-periodic condition in the imaginary time $t$ on the spinor field,  
\begin{align}\label{temporal periodicity}
    \psi_j(t,\boldsymbol{r}) = -\psi_j(t+\beta,\boldsymbol{r}) ,
\end{align}
one can prove that the allowed frequencies must have the form
\begin{align}\label{omega n}
    \omega_n = \frac{2\pi}{\beta}\left(n + \frac{1}{2}  \right) , \quad \forall~n \in \mathbb{Z} . 
\end{align}
The condition \eqref{temporal periodicity} corresponds to compacting the imaginary-time dimension $t$ into a circumference of length $\beta$. This amounts to considering spinor fields defined over a four-dimensional space with topology of the type $\mathbb{S}^1 \times \mathbb{R}^{3}$, where periodicity represented by $\mathbb{S}^1$ is oriented at the $t$-direction. 
In Refs. \cite{Kulikov1989, Ahmadi2005,Joas2017}, spinor fields are worked out in several spaces whose topology is formed from the direct products.

Because of the time decomposition \eqref{time decomposition}, it is particularly useful to write the operator $\slashed{D}^2(m)$ as
\begin{equation}\label{Laplace type operator}
        \slashed{D}^2(m) =  L_1 + \slashed{\nabla}^2(m) ,
\end{equation}
where $L_1$ and $\slashed{\nabla}^2(m)$ are defined as follows
\begin{align}
    L_1 = - \partial^{2}_{t} \quad \text{and} \quad \slashed{\nabla}^2(m) = -\partial^i \partial_{i} + m^2. 
\end{align}
$\slashed{\nabla}^2(m)$ is an elliptic, self-adjoint, second-order differential spinor operator defined on the spatial part of $M$. The generalized zeta function method associated with the scalar operators defined on spaces with different conditions can be found in Refs. \cite{Hawking1977,Giulia2021}.
The trace of the operator $\slashed{D}^{2}(m)$ can also be split into temporal and spatial parts through the trace property
\begin{align}\label{Trace property}
    \text{Tr}\left[e^{-\tau \slashed{D}^{2}(m)}\right]  = 4\,\text{Tr}\left(e^{-\tau L_1 }\right) \text{Tr}\left[e^{-\tau \slashed{\nabla}^2(m)} \right] ,
\end{align} 
where the multiplicative factor $4$ is due to the spinor nature of $\slashed{D}^{2}(m)$.  
The eigenvalues of $L_1 = -\partial_t^{2}$ can be obtained from Eq. \eqref{omega n}, so we have that
\begin{align}\label{Trace of del t}
\text{Tr}\left(e^{-\tau L_1}\right) =  \sum_{n= - \infty}^{\infty} e^{-\tau \frac{4\pi^2}{\beta^2} \left(n + \frac{1}{2}\right)^2} .
\end{align}
Defining the constant parameters ${a = 4\pi^2 \tau/\beta^2}, b = n$ and $c = 1/2$, and using the Jacobi inversion identity \cite{Kirstein2010},
\begin{align}
    \sum_{n= - \infty}^{\infty} e^{- a\left(b + c\right)^2} = \sqrt{\dfrac{\pi}{a}}  \sum_{n= - \infty}^{\infty} e^{- \frac{\pi^2}{a} b^2 - 2 \pi i b c } ,
\end{align}
we can rewrite Eq. \eqref{Trace of del t} as follows:
\begin{align}\label{Eigenvalues of L1}
\text{Tr}\left(e^{-\tau L_1}\right) = \dfrac{\beta}{\sqrt{4\pi \tau}} \left[1 +  2\sum_{n= 1}^{\infty} \cos(\pi n) e^{-\frac{\beta^2}{4 \tau}n^2} \right] ,
\end{align}
in which the first term inside the brackets represents \text{the ${n = 0}$ term in the series}. 
Summing up these results, one eventually obtains the integral representation of the zeta function $\zeta_{\slashed{D}^2(m)}$ associated with $\slashed{D}^2(m)$, which is a Laplace type operator defined in a flat space with a metric of Euclidean signature and acts on a spinor field in thermal equilibrium at finite temperature ${T = \beta^{-1}}$, satisfying anti-periodicity conditions. It follows from \eqref{Trace property}, \eqref{Eigenvalues of L1}, and \eqref{integral representation} that the zeta function $\zeta_{\slashed{D}^2(m)}$ can be put in the form
\begin{align}\label{Zeta of D(m) square}
   &{~}\zeta_{\slashed{D}^{2}(m)}(z)   =   \frac{\beta}{\sqrt{4\pi}} \left[  \frac{ Z_1(z)}{\Gamma(z)}  +  Z_2(z, \beta) \right] ,
\end{align}
with
\begin{align}
    Z_1(z) &= \,\Gamma\left(z- 1/2\right) \zeta_{\slashed{\nabla}^2(m)}\left(z- 1/2\right), \label{Z1} \\
    Z_2(z,\beta)& = \frac{2}{\Gamma(z)} \sum_{n=1}^{\infty} \cos(\pi n) \times \nonumber\\
    &\times \int_{0}^{\infty} d\tau\,\tau^{z-\frac{3}{2}} e^{-\frac{\beta^2}{4 \tau}n^2}  K_{\slashed{\nabla}^2(m)}(\tau) \label{Z2}. 
\end{align}
where $\zeta_{\slashed{\nabla}^2(m)}$ and $K_{\slashed{\nabla}^2(m)}$ are the zeta function and the global heat kernel associated with the spinor operator $\slashed{\nabla}^2(m)$. 

Once the zeta function is obtained, we should compute the vacuum-free energy, Eq. \eqref{free energy}, which may have divergent parts. In order to analyze such divergences, it is convenient to perform the small-$\tau$ asymptotics expansion of the heat kernel 
\cite{Bordag2000}: 
\begin{align}\label{small t expansion}
    K_{\slashed{\nabla}^2(m)}(\tau) \sim \frac{e^{-\tau m^2}}{(4 \pi \tau)^{3/2}}\sum_{p = 0, 1/2, 1,\ldots } c_{p}\,\tau^{p} ,
\end{align}
where $c_p = c_p(\slashed{\nabla}^2)$ are the heat kernel coefficients associated with the massless operator $\slashed{\nabla}^2$. 
Now, using the integral representation of $\zeta_{\slashed{\nabla}^2(m)}$, Eq. \eqref{integral representation}, we can use this asymptotic behavior of $K_{\slashed{\nabla}^2(m)}$ to write the function $Z_1(z)$ as
\begin{align}\label{Gamma times zeta}
  Z_1(z)  &= \int_{0}^{\infty} d\tau\,\tau^{z-3/2} K_{\slashed{\nabla}^2(m)}(\tau)\nonumber\\
      & =  \frac{1}{(4\pi)^{3/2}}\sum_{p = 0, 1/2, 1,\ldots } \frac{c_p\,\Gamma\left(z + p - 2 \right)}{ \left(m^2\right)^{z + p - 2}} , 
\end{align}
which has simple poles located at 
\begin{align}\label{poles constraint}
    z + p - 2 = - \kappa, \quad \forall~\kappa \in \mathbb{N},   
\end{align}
since the gamma function diverges only at non-positive integers, with the corresponding residues containing  non-negative mass exponents
\begin{align}
    \text{Res}\left(Z_1(z), -\kappa \right) =  \frac{ (-1)^\kappa\,c_{2-\kappa - z} \, m^{2\kappa}}{(4\pi)^{3/2} \kappa!}. 
\end{align}
As we are only interested in the limit $z \rightarrow 0$, the constraint \eqref{poles constraint} translates into considering the series \eqref{Gamma times zeta} up to order $p \leq 2$, to be consistent with the poles at $\kappa = 0, 1, 2$. In particular, this means that the terms in the series with semi-integer $p$ have no poles, the divergent contributions come from the dominant coefficients 
$c_0, c_1$ and $c_2$, with $c_0$ and $c_1$ multiplied by non-negative mass exponents. However, these divergent contributions are canceled out by the pole in $\Gamma(z)$ in the denominator of $\zeta_{\slashed{D}^{2}(m)}(z)$. Indeed, near $z=0$
\begin{align}
    \frac{1}{\Gamma(z)} = z + \gamma_{E} z^2 + \mathcal{O}(z^3), 
\end{align}
where $\gamma_{E}$ is the Euler constant. In particular, this implies that $Z_{2}(0,\beta) = 0$, since the remaining integral in Eq. \eqref{Z2} is finite at $z = 0$. Thus,
\begin{align}\label{Zeta of D(m) square at zero}
   &{~}\zeta_{\slashed{D}^{2}(m)}(0)   =  \frac{\beta}{16\pi^2}\,c_2(m) , 
\end{align}
where
\begin{align}\label{c_2(m) heat kernel coefficient}
    c_2(m) = \sum_{\kappa=0}^{2} \frac{(-1)^\kappa}{\kappa!} c_{2-\kappa} m^{2\kappa} = \frac{m^4 c_0}{2} - m^2 c_1 + c_2 .
\end{align}


In order to obtain the expression for $\zeta'_{\slashed{D}^{2}(m)}(0)$, we should note that while $Z_2(z, \beta) $ and its first derivative with respect to $z$ are finite at $z=0$, $Z_1(z)$ has a pole at $z=0$ coming from the pole of ${\zeta_{\slashed{\nabla}^2(m)}\left(z - 1/2\right)}$  at this point,  with residue 
\begin{align}
     \text{Res}(\zeta_{\slashed{\nabla}^2(m)}\left(z - 1/2\right), z = 0) =  -\frac{c_2(m)}{16\pi^2} .
\end{align}
So, separating off this pole contribution and taking the derivative of $\zeta_{\slashed{D}^{2}(m)}(z)$ with respect to $z$, after some algebra, leads to the relation for the regularized (reg) free energy for the Dirac field as follows
\begin{align}\label{Reg free energy}
    &F(\beta, m, \mu) = -\frac{1}{2}\,\text{FP}[\zeta_{\slashed{\nabla}^2(m)}(-1/2)] \nonumber\\
    &+ \frac{c_2(m)}{16\pi^2}\left\{ \text{ln}\left(\mu^2\right) + 2[1- \text{ln}(2)]  \right\} \nonumber\\
    &+ \frac{1}{\sqrt{4\pi}} \sum_{n= 1}^{\infty} \cos(\pi n)\int_{0}^{\infty} d\tau\,\tau^{-3/2}  e^{-\frac{\beta^2}{4 \tau}n^2} K_{\slashed{\nabla}^2(m)}(\tau) , 
\end{align}
where $\text{FP}[\zeta_{\slashed{\nabla}^2(m)}(-1/2)]$ stands for the finite part of $\zeta_{\slashed{\nabla}^2(m)}(-1/2)$.  This result corresponds to the massive spinor counterpart of the one obtained by Kirstein in Ref. \cite{Kirstein2010} for the massless scalar field, in which the only nonvanishing heat kernel coefficient is $c_2$.  Finally, taking the limit $\beta \rightarrow \infty$, we obtain the following expression for zero-temperature free energy associated with the massive spinor field
\begin{align}\label{Effetive energy}
   E(m, \mu) &= \underset{\beta \rightarrow \infty}{\text{lim}} F(\beta, m, \mu) \nonumber\\
   & = -\frac{1}{2}\,\text{FP}[\zeta_{\slashed{\nabla}^2(m)}(-1/2)] + \frac{c_2(m)}{16\pi^2}\,\text{ln}(\tilde{\mu}^2), 
\end{align}
where the rescaled parameter $\tilde{\mu} = \mu e/2$ has been employed.
The $\beta$-dependent remaining term in $\text{Eq.~\eqref{Reg free energy}}$ is the temperature correction to the free energy given by
\begin{align}\label{Generic temperature correction}
  \Delta F(\beta, m) & =  \frac{1}{\sqrt{4\pi}} \sum_{n= 1}^{\infty} \cos(\pi n) \times \nonumber\\
  & \times \int_{0}^{\infty} d\tau\,\tau^{-3/2}  e^{-\frac{\beta^2}{4 \tau}n^2} K_{\slashed{\nabla}^2(m)}(\tau) . 
\end{align}
At this stage, it is worth noting that there remains no singularity when $z \rightarrow 0$. So, the spinor free energy is finite. 
However, when the heat kernel coefficient $c_2(m)$ is nonvanishing, the zeta function regularization prescription becomes ambiguous due to its natural dependence on the arbitrary parameter $\mu$, which has been rescaled without loss of generality. This scale freedom when $c_2(m) \neq 0$ is also responsible for the so-called conformal anomaly \cite{Vassilevich2011, Vassilevich2003}. It is worth mentioning that in the massless case ($m=0$), all information concerning free energy ambiguity is contained in the $c_2$ coefficient so that such ambiguity is present only if $c_2 \neq 0$. 

To ensure the uniqueness of the renormalization process, such ambiguity can be removed by the subtraction of the contribution arising from the heat kernel coefficients $c_p$ with $p \leq 2$. After performing this finite renormalization, the remaining part can be expressed as the sum of the zero-temperature Casimir energy $E_{\text{cas}}(m)$ plus the temperature correction $\Delta F(\beta, m)$
\begin{align}\label{Renormalized free energy}
    F(\beta, m) = E_{\text{cas}}(m) + \Delta F(\beta, m), 
\end{align}
where the Casimir energy at zero temperature $E_{\text{cas}}(m)$ is as follows
\begin{align}\label{Renormalized Casimir energy}
   E_{\text{cas}}(m) = -\frac{1}{2}\,\text{FP}[\zeta_{\slashed{\nabla}^2(m)}(-1/2)]. 
\end{align}
Note that as the Casimir energy exhibits a mass dependence of the type $e^{-\tau m^2}$, serving as a convergence factor in the integral representation of $\zeta_{\slashed{\nabla}^2(m)}(-1/2)$, it must vanish when the mass tends to infinity. This is due to the fact that there can be no quantum fluctuations at this limit.


In the same fashion, one can use the small-$\tau$ heat kernel expansion, which can also be seen as a {\text{large-}$m$} expansion, to fix the ambiguity problem uniquely. Since the heat kernel coefficient $c_2(m)$ increases with non-negative powers of the mass, one must require that $E(m, \mu)$ should be renormalized to zero for large $m$ \cite{Bordag2000}
\begin{align}
    \underset{m \rightarrow \infty}{\text{lim}} E(m, \mu)  \rightarrow 0 ,
\end{align}
removing all the dependence on the scale factor $\mu$ of the Casimir energy. It is worth pointing out that when $c_2(m)$ is identically null, i.e., when the FP prescription is redundant, the finite renormalization is unnecessary because the ambiguity is naturally removed, and hence the scale freedom is broken. 

So far, the zeta function method has been utilized to obtain a generic expression for the vacuum free energy associated with a massive spinor field defined on a four-dimensional flat space endowed with a Euclidean metric. In particular, given the anti-periodic condition of spinor fields in imaginary time, we have been able to find a constraint on the eigenvalues of $L_1$. From now on, we shall consider a space with a circularly compactified dimension that imposes an anti-periodic boundary condition along a vector on the spinor field. By imposing this spatial condition, we will explicitly obtain the restrictions that the eigenvalues of $\slashed{\nabla}^2(m)$ must obey, hence evaluating the spinor vacuum free energy.

\section{Spinor field in a nontrivial compactified space}\label{Helix topology}


This section aims to find an analytical expression for the zero-temperature Casimir energy and its corresponding temperature corrections induced by a topological constraint simulating a boundary condition imposed on the spinor field along a vector in plane. To accomplish this, we will adopt the heat kernel approach to zeta-function regularization.

Consider the space $M = (\mathbb{R}^4, \boldsymbol{g})$, where $\boldsymbol{g}$ is a positive-definite symmetric metric whose components are given by Eq. \eqref{metric component}, namely $\delta_{ab}$, so that the line element on $M$ takes the form
\begin{align}
  ds^2 = dt^2 + dx^2 + dy^2 +dz^2 ,
\end{align}
where $\{x^a \} = \{t, x, y, z\}$ are cartesian coordinates. We recall that the coordinate $t$ is compactified into a circumference length $\beta$ as discussed in \text{Sec. \ref{Path integral}}, equivalent to equipping $M$ with the topology $\mathbb{S}^1(\text{time}) \times \mathbb{R}^3$.  
Here we consider  the space $\mathbb{R}^3$ with a circularly compactified dimension, where the periodicity represented by the circle $\mathbb{S}^1$ is oriented in the direction of a vector $\boldsymbol{L} \in \mathbb{R}^{2}$ given by
\begin{align}\label{compact vector}
    \boldsymbol{L} = a\,\boldsymbol{e}_{x} - b \,\boldsymbol{e}_{y} , \quad  ~a, b \in \mathbb{R},
\end{align}
referred to here as compact vector. $\boldsymbol{e}_{x}$ and $\boldsymbol{e}_{y}$ denote the unit vectors along the directions $x$ and $y$, respectively, and the parameters $a$ and $b$ constant displacements. 
In particular, the compact dimension size is determined by the vector length
\begin{align}
  L(a, b) = |\boldsymbol{L}| = \sqrt{a^2 + b^2}  .
\end{align} 
Although not along a coordinate axis as usual, the compactification in a $\mathbb{S}^1$ topology along $\boldsymbol{L}$ is quite natural since $L$ is a homogeneous function of degree $1$, that is ${L(n a, n b) = n\,L(a, b)}$ for all non-null integer $n$. Choosing a suitable frame field can recover the usual $\mathbb{S}^1$ topology, as we will see later. 

Along the compact vector, the spinor field is assumed to satisfy the following boundary condition
\begin{align}\label{Helix condition}
    \psi_j(t, \boldsymbol{r}) = - \psi_j(t, \boldsymbol{r} + \boldsymbol{L}),
\end{align}
similar to the temporal anti-periodicity condition, $\text{Eq. \eqref{temporal periodicity}}$. In fact, when  ${b=0}$ and ${a \neq 0}$, the spinor field satisfies a spatial anti-periodicity condition, ${ \psi_j(t, x, y, z) = - \psi_j(t, x + a, y, z)}$, induced by the compact subspace of the coordinate $x$. The condition \eqref{Helix condition} means that the spinor field undergoes a sign change after traveling a distance $a$ in the $x$-direction and $b$ in the $y$-direction and returns to its initial value after traveling distances $2a$ and $2b$, namely ${\psi_j(t, \boldsymbol{r}) =  \psi_j(t, \boldsymbol{r} + 2\boldsymbol{L})}$. In particular, through a coordinate origin redefinition, without changing the orientation of the axes, one can equally write the condition \eqref{Helix condition} as  ${\psi_j(t, x+a, y, z) = - \psi_j(t, x, y+b, z)}$. In Ref.  \cite{Xiang2011}, this latter condition was investigated in the helix-like topology context and called the anti-helix condition, with $a$ and $b$ labeling the circumference length and pitch of the helix, respectively.


An ansatz for the massive spinor field in the geometry of the space $M$ was given in Eq. \eqref{time decomposition}, namely ${\psi_j(t, \boldsymbol{r}) = e^{-i\omega_n t} \chi_\ell(\boldsymbol{r})}$, with the spatial part $\chi_\ell(\boldsymbol{r})$ satisfying the eigenvalue equation
\begin{align}\label{spatial spinor field}
    \slashed{\nabla}^2(m) \chi_{\ell}(\boldsymbol{r}) = \left(\lambda_{\ell}^2 + m^2 \right) \chi_{\ell}(\boldsymbol{r}) .
\end{align}
The eigenfunctions $\chi_{\ell}$ 
of the above equation have the form
 \begin{align}\label{spatial spinor solution}
\chi_{\ell}(\boldsymbol{r}) = \mathcal{N}\, e^{i\,\boldsymbol{k} \cdot \boldsymbol{r}}\, u_{s}(\boldsymbol{k}),
 \end{align}
with $\mathcal{N}$ being a normalization constant and $u_{s}(\boldsymbol{k})$ being four-component spinors whose explicit form is unnecessary for our purposes. 
There are four spinors for each choice of momentum $\boldsymbol{k}$, two of which have positive energy and two with negative energy \cite{Xiang2011}.


We are interested in obtaining the finite temperature Casimir energy under the influence of the boundary condition \eqref{Helix condition}, which imposes the following non-trivial relation for the momentum along the compact vector 
\begin{align}\label{kx ky relation}
\boldsymbol{k} \cdot \boldsymbol{L} =  k_x a -  k_y b = 2\pi\left(n+\frac{1}{2}\right),  \quad \forall~ n \in \mathbb{Z}.
\end{align}
This means that the label $\ell$ in the spinor field \eqref{spatial spinor solution} should be understood as the set of quantum numbers ${\{\ell\} = \{n, k_y, k_z, s\}}$, since $k_x$ can be eliminated employing Eq. \eqref{kx ky relation}. In particular, the sum over $\ell$ becomes
\begin{align}
    \sum_{\ell}  \rightarrow \sum_{n } \int dk_y\int dk_z  \sum_{s} . 
\end{align}
Thus, utilizing the identification mentioned above in the completeness relation \eqref{orthonormality matricial} for the spinor field $\chi_\ell$ obeying the boundary condition \eqref{Helix condition}, we are left with the normalization constant
\begin{align}\label{normalization constant}
    \mathcal{N} = \frac{1}{2\pi\sqrt{a}} .
\end{align}

Given the spinor fields \eqref{spatial spinor solution}, one can determine the eigenvalues in Eq. \eqref{spatial spinor field}, allowing for the construction of the spinor heat kernel. Assuming that the requirement \eqref{kx ky relation} holds, the corresponding eigenvalues are found to be
\begin{align}\label{spatial eigenvalue}
    \lambda_{\ell}^2 & = k_x^2 + k_y^2 + k_z^2 \nonumber\\
    &= \left[ \frac{2\pi}{a}\left(n + \frac{1}{2}\right) + \frac{b}{a}\, k_y \right]^2 + k_y^2 + k_z^2 .
\end{align}

It is worth mentioning that with an appropriate choice of frame field, it is possible to align the compactification on $\mathbb{S}^1$ along one of the coordinate axes. In the momentum space, this can be achieved by defining for instance
\begin{align}
    k_y = \frac{a}{L}\left[k_{Y} + \frac{2\pi b}{a\,L}\left(n + \frac{1}{2}\right) \right] . 
\end{align}
This transformation leads to the eigenvalues \eqref{spatial eigenvalue} to be written as
\begin{align}
     \lambda_{\ell}^2 = \left[ \frac{2\pi}{L}\left(n + \frac{1}{2}\right)\right]^2 + k_{Y}^2 + k_z^2 .
\end{align}
These eigenvalues stem in particular from the spatial anti-periodic boundary condition ${\psi_j(t, X, Y, z) = - \psi_j(t, X+L, Y, z)}$ induced by the usual topology ${\mathbb{S}^1}(\text{space}) \times \mathbb{R}^2$, whereby the coordinate $X$ is compactified into a circumference length $L$.
In fact, along this compact dimension, the latter condition produces the discrete momentum  ${k_{X} = 2\pi (n + 1/2)/L}$. In particular, this means that in the limiting case ${b = 0}$, our results recover the ones presented in \text{Ref.~\cite{Bellucci2009}} for a specific case and include temperature corrections. In this study, the authors investigated the Casimir effect for spinor fields in toroidally compactified spaces, including general phases in the boundary condition along the compact dimensions. 

Building upon the previous results, we can introduce the heat kernel approach to obtain a zeta-function analytical expression for a spinor field defined on $M$ with the eigenvalues \eqref{spatial eigenvalue}.
Instead of the global heat kernel, it is more appropriate to utilize the local heat kernel. The reason is that the heat kernel carries information concerning the space where the field is defined, making it particularly valuable when focusing on the the influence of topological constraint imposed by the boundary conditions on the thermal vacuum fluctuations. 


\subsection{Spinor heat kernel and Casimir energy density}

The spinor heat kernel $K_{\slashed{\nabla}^2(m)}(\boldsymbol{r}, \boldsymbol{r}', \tau)$ is a two-point function locally defined as solutions of the heat conduction equation
\begin{equation}\label{heat equation}
    \left[\frac{\partial}{\partial \tau} +\slashed{\nabla}^2(m) \right]\,K_{\slashed{\nabla}^2(m)}(\boldsymbol{r}, \boldsymbol{r}', \tau) = 0  \quad \text{for}~\tau>0,
\end{equation}
supplemented with the initial condition
\begin{equation}\label{initial condition}
\lim_{\tau \rightarrow 0} K_{\slashed{\nabla}^2(m)}(\boldsymbol{r}, \boldsymbol{r}', \tau) = \delta(\boldsymbol{r} - \boldsymbol{r}') \,\mathbb{1}. 
\end{equation}
The operator $\slashed{\nabla}^2(m)$ is taken to act on the first argument of $K_{\slashed{\nabla}^2(m)}$. Similar to $\slashed{\nabla}^2(m)$, $K_{\slashed{\nabla}^2(m)}$ is represented by a $4 \times 4$ matrix. 

Taking into account Eq. \eqref{Eigenvalue equation fo D^2}, the solutions of \text{Eq. \eqref{heat equation}} can be expressed in terms of the eigenvalues and eigenfunctions of $\slashed{\nabla}^2(m)$
\begin{equation}\label{spinor heat kernel}
K_{\slashed{\nabla}^2(m)}(\boldsymbol{r}, \boldsymbol{r}', \tau) = e^{-\tau m^2}\sum_{\ell} e^{-\lambda_{\ell}^2 \tau} \chi_{\ell}(\boldsymbol{r}) \chi_{\ell}^{\dagger}(\boldsymbol{r}'). 
\end{equation}
One can verify that the above expression provides a solution to Eq. \eqref{heat equation}, as well as satisfying the initial condition \eqref{initial condition} since the spinor field obeys Eq. \eqref{orthonormality matricial}. 

Inserting the spinor solution \eqref{spatial spinor solution} along with the normalization constant \eqref{normalization constant} and eigenvalues \eqref{spatial eigenvalue} into spinor heat kernel \eqref{spinor heat kernel}, it follows the expression
\begin{align}\hspace{-1mm}
        K_{\slashed{\nabla}^2(m)}(\boldsymbol{r}, \boldsymbol{r}', \tau) &= \dfrac{e^{-\tau m^2}}{4 \pi^2 a }\nonumber\\ 
        &\times \sum_{n}  e^{-\frac{4\pi^2\tau}{a^2} \left(n +\frac{1}{2}\right)^2 + \frac{2\pi i}{a} \left(n +\frac{1}{2}\right)\Delta x} \nonumber\\
        &\times\int dk_y \, e^{-\tau \frac{L^2}{a^2} k_y^2 + \left[i \Delta v -\frac{4 \pi b \tau }{a^2} \left(n + \frac{1}{2}\right) \right]  k_y}\nonumber\\
        &\times\int dk_z e^{-\tau k_z^2 + i k_z  \Delta z}\,\mathbb{1}, \label{HK}
\end{align}
where
\begin{align}\label{gamma}
  \Delta v = \frac{b}{a}\, \Delta x + \Delta y . 
\end{align}
We can write Eq. \eqref{HK} in a more compact form. To perform this, let us define the complex parameters $w$ and $q$ as follows
\begin{align}
    w =  \frac{b\Delta v}{L^2} - \frac{\Delta x}{a} - \frac{q}{2}\quad \text{and} \quad q =  \frac{4 \pi i \tau}{L^2} ,
\end{align}
and introduce the following Jacobi function defined in terms of the parameters $w$ and $q$ as \cite{Elizalde1994}
\begin{align}
    \theta_3(w, q) = \sum_{n=-\infty}^{\infty} e^{i\pi q n^2 - 2\pi i w n} . 
\end{align}
Evaluating the integrals over the independent momenta $k_y$ and $k_z$ in Eq. \eqref{HK}, we end up with the following relation between the spinor heat kernel and the Jacobi function
\begin{align}
    K_{\slashed{\nabla}^2(m)}(\boldsymbol{r}, \boldsymbol{r}', \tau) = \frac{e^{-\frac{\left|\boldsymbol{r} - \boldsymbol{r}'\right|^2}{4\tau} -\tau m^2} }{\left( 4\pi \tau \right)^{3/2}} \frac{\sqrt{-i q}}{e^{-i\pi \frac{\omega^2}{q}}}\, \theta_3(w, p)\, \mathbb{1} .
\end{align}
Since we are interested in the contributions coming from the topology for the thermal vacuum fluctuations, it is convenient to separate the Euclidean part of the heat kernel, which should not depend on the topology parameters. This can be done by rewriting the $\theta_3$ Jacobi function utilizing the following identity
\begin{align}
    \theta_3(w, q) =  \frac{1}{\sqrt{-i q}}\,e^{-i\pi \frac{\omega^2}{q}} \, \theta_3\left(\frac{w}{p}, -\frac{1}{q}\right) . 
\end{align}
Employing this identity, leads to the following expression
\begin{align}\label{HK1}
    K_{\slashed{\nabla}^2(m)}(\boldsymbol{r}, \boldsymbol{r}', \tau) = K_{\slashed{\text{E}}}(\boldsymbol{r}, \boldsymbol{r}', \tau) \sum_{n = -\infty}^{\infty} e^{-\frac{L^2}{4\tau} n^2 + i\pi n } ,
\end{align}
with
\begin{align}\label{Euclidean heat kernel function}
   K_{\slashed{\text{E}}}(\boldsymbol{r}, \boldsymbol{r}', \tau) = \frac{1}{\left( 4\pi \tau \right)^{3/2}} \, e^{-\frac{\left|\boldsymbol{r} - \boldsymbol{r}'\right|^2}{4\tau} -\tau m^2}\, \mathbb{1} .
\end{align}
where $K_{\slashed{\text{E}}}$ is the spinor version of the well-known Euclidean heat kernel associated with the massive scalar Laplacian operator defined on the flat space $\mathbb{R}^{3}$ \cite{Vassilevich2003}.
Note that $K_{\slashed{\text{E}}}$ is identified with the term $n=0$ in the series.

Let us now explore the heat kernel properties at the coincidence limit $\boldsymbol{r}'\to \boldsymbol{r}$ in Eq. \eqref{HK1} which results in
\begin{align}\label{General heat kernel}
   K_{\slashed{\nabla}^2(m)}(\boldsymbol{r}, \boldsymbol{r}, \tau) &=  K_{\slashed{\text{E}}}(\boldsymbol{r}, \boldsymbol{r}, \tau) \nonumber\\
    & +  \frac{e^{-\tau m^2}}{(4\pi \tau)^{3/2} }\sum_{n = 1}^{\infty} 2\cos(\pi n) e^{-\frac{L^2}{4\tau} n^2} \, \mathbb{1}  .
\end{align} 
The first term on the right-hand side corresponds to the Euclidean heat kernel, while the second term encodes information about the space topology, as can be seen from its dependence on parameter $L$. 
For small $\tau$, the heat kernel admits an expansion in powers of $\tau$, with coefficients reflecting the space configuration. In our case, by evaluating the above series at small $\tau$, one can see that all terms are exponentially small except for the one associated with the Euclidean heat kernel contribution (${n=0}$). Thus, the spinor heat kernel $K_{\slashed{\nabla}^2(m)}(\boldsymbol{r})$ on Euclidean geometry with a circular compactification along $\boldsymbol{L}$ exhibits an asymptotic behavior similar to the one considered in Eq. \eqref{small t expansion} with only one non-vanishing heat kernel coefficient
\begin{align}\label{Local heat kernel expansion}
 K_{\slashed{\nabla}^2(m)}(\boldsymbol{r}, \boldsymbol{r}, \tau)  \sim   \frac{e^{-\tau m^2}}{\left( 4\pi \tau \right)^{3/2}}  \sum_{p} c_p(\boldsymbol{r}) \tau^p + \mathcal{O}(e^{-1/\tau})  , 
\end{align}
where the local heat kernel coefficients $c_p(\boldsymbol{r})$ are given by
\begin{align}
    c_p(\boldsymbol{r}) = \delta_{0p} \,\mathbb{1} \quad \forall ~p .
\end{align}
$\mathcal{O}(e^{-1/\tau})$ stands for those terms going to zero faster than any positive power of $\tau$ and, therefore, can be neglected. 
In contrast with the global case, the local heat kernel coefficients carry spinor indices, hence $4 \times 4$ matrices.   
Note that $K_{\slashed{\nabla}^2(m)}(\tau)$ can be obtained from $K_{\slashed{\nabla}^2(m)}(\boldsymbol{r}, \boldsymbol{r}', \tau)$ performing an integral in the whole space
\begin{align}
   K_{\slashed{\nabla}^2(m)}(\tau) = \int d^3\boldsymbol{r}\,\sqrt{g^{(3)}}\,\text{tr}\left[K_{\slashed{\nabla}^2(m)}(\boldsymbol{r}, \boldsymbol{r}, \tau)\right], 
\end{align}
where $g^{(3)}$ is the spatial part of the metric determinant, and the trace operation tr is taken over the spinor indices only. Thus,  the global heat kernel coefficient $c_0$ is found to be
\begin{align}\label{c0 heat kernel coefficient}
    c_0 & = 4\, V_3 ,
\end{align}
where $V_3$ is the volume of the $3$-dimensional base space of $M$. 
As discussed in Sec. \ref{Zeta function of an operator}, for nonvanishing heat kernel coefficients $c_p \,(p \leq 2)$, the zeta function is not finite, and the renormalization procedure is not unique. In fact, although the vacuum energy is finite due to the FP prescription introduced in the Casimir energy, the coefficient $c_0$ gives origin to the terms in the vacuum energy, which increase with non-negative powers of the mass, besides the logarithmic dependence on the scale factor $\mu$. To ensure a unique renormalization procedure and obtain an unambiguous spinor vacuum free energy, all contributions associated with $c_0$ should be disregarded, thereby renormalizing the energy to zero for large masses.

After performing the finite renormalization, we can proceed with the analytical calculation of the spinor vacuum free energy. 
First, we should note that even though we are working in the local regime, the two-point function $K_{\slashed{\nabla}^2(m)}(\boldsymbol{r}, \boldsymbol{r}, \tau)$ is coordinate-independent.  Therefore, considering that global quantities can be derived from local ones by integrating over the space coordinates, the local version of the spinor vacuum free energy differs from the global version by a volume element and retains the same form as Eq. \eqref{Renormalized free energy}, namely
\begin{align}\label{Renormalized free energy density}
    \mathcal{F}(\beta, m) = \mathcal{E}_{\text{cas}}(m) + \Delta \mathcal{F}(\beta, m), 
\end{align}
where $\mathcal{E}_{\text{cas}}(m)$ is then the Casimir energy density at zero temperature
\begin{align}\label{Renormalized Casimir energy density}
   \mathcal{E}_{\text{cas}}(m) = -\frac{1}{2}\,\text{FP}[\zeta_{\slashed{\nabla}^2(m)}(\boldsymbol{r}, -1/2)],
\end{align}
and $\Delta \mathcal{F}(\beta, m)$ is the temperature correction with the form
\begin{align}\label{Temperature corrections density}
  \Delta &\mathcal{F}(\beta, m)  =  \frac{1}{\sqrt{4\pi}} \sum_{n= 1}^{\infty} \cos(\pi n) \times \nonumber\\
  & \times \int_{0}^{\infty} d\tau\,\tau^{-3/2}  e^{-\frac{\beta^2}{4 \tau}n^2} \text{tr}\left[ K_{\slashed{\nabla}^2(m)}(\boldsymbol{r}, \boldsymbol{r}, \tau) \right] . 
\end{align}
Here, the trace operation tr is taken over the spinor indices only and the local zeta function is defined in terms of $K_{\slashed{\nabla}^2(m)}(\boldsymbol{r}, \boldsymbol{r}, \tau) $ giving rise to
\begin{align}\label{Local zeta function}
  \zeta_{\slashed{\nabla}^2(m)}&(\boldsymbol{r}, z-1/2) =  \frac{1}{\Gamma(z-1/2)}\nonumber\\
&\times   \int_{0}^{\infty} d\tau\,\tau^{z-3/2} \, \text{tr}\left[ K_{\slashed{\nabla}^2(m)}(\boldsymbol{r}, \boldsymbol{r}, \tau) \right] . 
\end{align}
Then, inserting the spinor heat kernel \eqref{General heat kernel} into Eq. \eqref{Local zeta function}, we conclude from Eq.  \eqref{Renormalized Casimir energy density} that the renormalized expression for the zero temperature Casimir energy density associated with a spinor field of mass $m$ depends on the topology parameter $L$ according to the relation
\begin{align}\label{Massive Casimir}
    \mathcal{E}_{\text{cas}}(m, L) = \frac{2}{\pi^2  L^4} \sum_{n=1}^{\infty} \frac{\cos(\pi n)}{n^2} (m L)^2 \, K_2(n m L) ,
\end{align}
where $K_{2}(z)$ is the MacDonald function. Note that the FP prescription removed the divergent contribution provided by the Euclidean heat kernel. The above result is exactly the one shown in Ref. \cite{Xiang2011} obtained in a different approach than the one presented here for both massive and massless spinors. 
In particular, the massless one can be obtained by making use of the following limit
\begin{align}\label{Small arguments limit}
    \underset{z \rightarrow 0}{\text{lim}}\,z^2 K_{2}(n z) = \frac{2}{n^2} . 
\end{align}
In fact, by separating the even and odd terms in the series \eqref{Massive Casimir}, and using the above equation, one can promptly verify that the following massless limit holds
\begin{align}
  \mathcal{E}_{\text{cas}}(L)  & = \underset{m \rightarrow 0}{\text{lim}} \mathcal{E}_{\text{cas}}(m, L) \nonumber\\
  & = \frac{1}{4\pi^2 L^4}\left[ \zeta(4) - \zeta\left(4, \frac{1}{2}\right)\right]   ,
\end{align}
where $\zeta(z)$ is the standard Riemann zeta function and $\zeta(z, w)$ is the Hurwitz zeta function defined for $\text{Re}(z)> 1$ and $w \neq 0, -1, -2, \ldots$, in the form \cite{Elizalde1994}
\begin{align}
    \zeta(z, w) = \sum_{n=0}^{\infty} (n + w)^{-z} .
\end{align}
Using the relation
\begin{align}
    \zeta\left(z, \frac{1}{2}\right) = (2^z -1) \,\zeta(z), 
\end{align}
along with the fact that ${\zeta(4) = \pi^4/90}$, we are left with the expression for the Casimir energy density, at zero temperature, associated with a massless spinor field
\begin{align}\label{Massless Casimir energy density}
    \mathcal{E}_{\text{cas}}(L) =  -\frac{7}{2}\,\frac{\pi^2}{90\,L^4} ,
\end{align}
It depends only on topology parameter, in complete agreement with the massless case obtained in Ref. \cite{Xiang2011}.  
In particular, its value is also $7/2$ times the result found in the massless scalar case under periodic boundary conditions along the compact vector $\boldsymbol{L}$ \cite{Zhai, Giulia2021}. It is worth mentioning that this case is unambiguous since the heat kernel coefficient $c_2$ is identically zero, so the renormalization procedure is unnecessary. 

 If one is interested in the limit $ m L  \gg 1$, then it is legitimate to consider  Mcdonald's function behavior at ${\text{large}~z}$
\begin{align}
    K_2(z) \simeq \left(\frac{\pi}{2 z}\right)^{1/2} e^{-z} \quad \text{for} \quad |\text{arg}(z)| < \pi/2. 
\end{align}
 In this limiting case, one can see that the Casimir energy density decays exponentially with the mass of the field
\begin{align}
    \mathcal{E}_{\text{cas}}\left(m  \gg \frac{1}{L}\right) = -\frac{2 m^2}{\pi^2 L^2 } \sqrt{\frac{\pi}{2 m L}} \, e^{- m L} ,
\end{align}
as expected, since an infinitely heavy field should not present quantum fluctuations and hence should not produce Casimir energy \cite{Bordag2000}. In Ref. \cite{Maluf2020}, a similar analysis is carried out for the Casimir energy for a real scalar field and the Elko neutral spinor field in a field theory at a Lifshitz fixed point.

\subsection{Finite-temperature corrections}

Let us now investigate the temperature correction, $\Delta \mathcal{F}(\beta, m)$, to the vacuum energy densities. Inserting the heat kernel \eqref{General heat kernel} into $\Delta \mathcal{F}(\beta, m)$ defined in Eq. \eqref{Temperature corrections density} leads to the following analytical expression for the temperature correction associated with the massive spinor field, in terms of a double sum 
\begin{align}
    \Delta \mathcal{F}(\beta, m, L) & =  \Delta \mathcal{F}_{\text{E}}(\beta, m) \nonumber\\
    & + \frac{4 m^4}{\pi^2} \sum_{n=1}^{\infty}\sum_{p=1}^{\infty} \cos(\pi n) \cos(\pi p) \nonumber\\
    & \times f_2\left(m \beta \sqrt{p^2 + \frac{L^2}{\beta^2} n^2}\right) .
\end{align}
For notational simplicity, we introduced the function $f_\nu(z)$ related to the Mcdonald function $K_{\nu}(z)$ as follows
\begin{align}
    f_\nu(z) = \frac{K_\nu(z)}{z^{\nu}} . 
\end{align}
The term $\Delta \mathcal{F}_{\text{E}}(\beta, m)$ is the contribution coming from the Euclidean heat kernel and thus does not depend on the parameter $L$. It has the following form
\begin{align}\label{Euclidean vacuum-free energy}
    \Delta \mathcal{F}_{\text{E}}(\beta, m) = \frac{2 m^4}{\pi^2} \sum_{n=1}^{\infty} \cos(\pi n) \, f_2(m\beta n) .
\end{align}
In particular, from Eq. \eqref{Small arguments limit}, we conclude that the temperature correction term in the massless limit is as follows
\begin{align}
  \Delta \mathcal{F}_{\text{E}}\left(\beta\right)  =  - \frac{7}{2} \frac{\pi^2}{90} \, T^4 ,
\end{align}
the standard black body radiation energy density associated with the massless spinor field.  
As we have seen, this contribution is directly related to the non-null coefficient $c_0$. In more general spaces with nontrivial topology, however, there may be temperature corrections to the above Stefan-Boltzmann law, proportional to $T^4$, coming from heat kernel coefficients associated with spacetime topology. These coefficients vanish in the limit of infinite space \cite{Herondy2023, Basil1978}.

Since the Casimir effect is a purely quantum phenomenon, the above term should not dominate in the high-temperature limit. Although not divergent, this quantum term should be subtracted in the renormalization procedure to obtain a correct classical contribution in this  limit. By doing so, we end up with the renormalized version of the free energy \eqref{Renormalized free energy density}
\begin{align}\label{Renormalized vacuum-free energy}
   & \mathcal{F}_{\text{ren}}(\beta, m, L) = \frac{2 m^4}{\pi^2} \sum_{n=1}^{\infty} \cos(\pi n) f_2\left(n m L\right) \nonumber\\ 
   &  + \frac{4 m^4}{\pi^2} \sum_{n=1}^{\infty}\sum_{p=1}^{\infty} \cos(\pi n) \cos(\pi p)  f_2\left(m \beta \sqrt{p^2 + \frac{L^2}{\beta^2} n^2}\right) .
\end{align}
In particular, by using the limit \eqref{Small arguments limit}, the above expression yields the massless contribution
\begin{align}\label{Massless free energy density}
  \mathcal{F}_{\text{ren}}(\beta, L) =  \mathcal{E}_{\text{cas}}(L) + \frac{8}{\pi^2 \beta^4} \sum_{n=1}^{\infty}\sum_{p=1}^{\infty} \frac{\cos(\pi n) \cos(\pi p) }{\left(p^2 + \frac{L^2}{\beta^2} n^2\right)^2 } , 
\end{align}
where $\mathcal{E}_{\text{cas}}(L)$ is the Casimir energy density associated with massless spinor field at zero temperature, Eq. \eqref{Massless Casimir energy density}. The presence of the double sum is convenient if one is interested in obtaining the low- and high-temperature asymptotic limits. Although the final result is the same, performing the sum in $p$ first is more straightforward for obtaining the high-temperature limit, while performing the sum in $n$ first is less complicated for obtaining the low-temperature limit. The final result for the free energy is equivalent. Choosing which one first is a simple question of convenience to attain our purposes.

\begin{figure}
    \centering
    \includegraphics[width=0.47\textwidth]{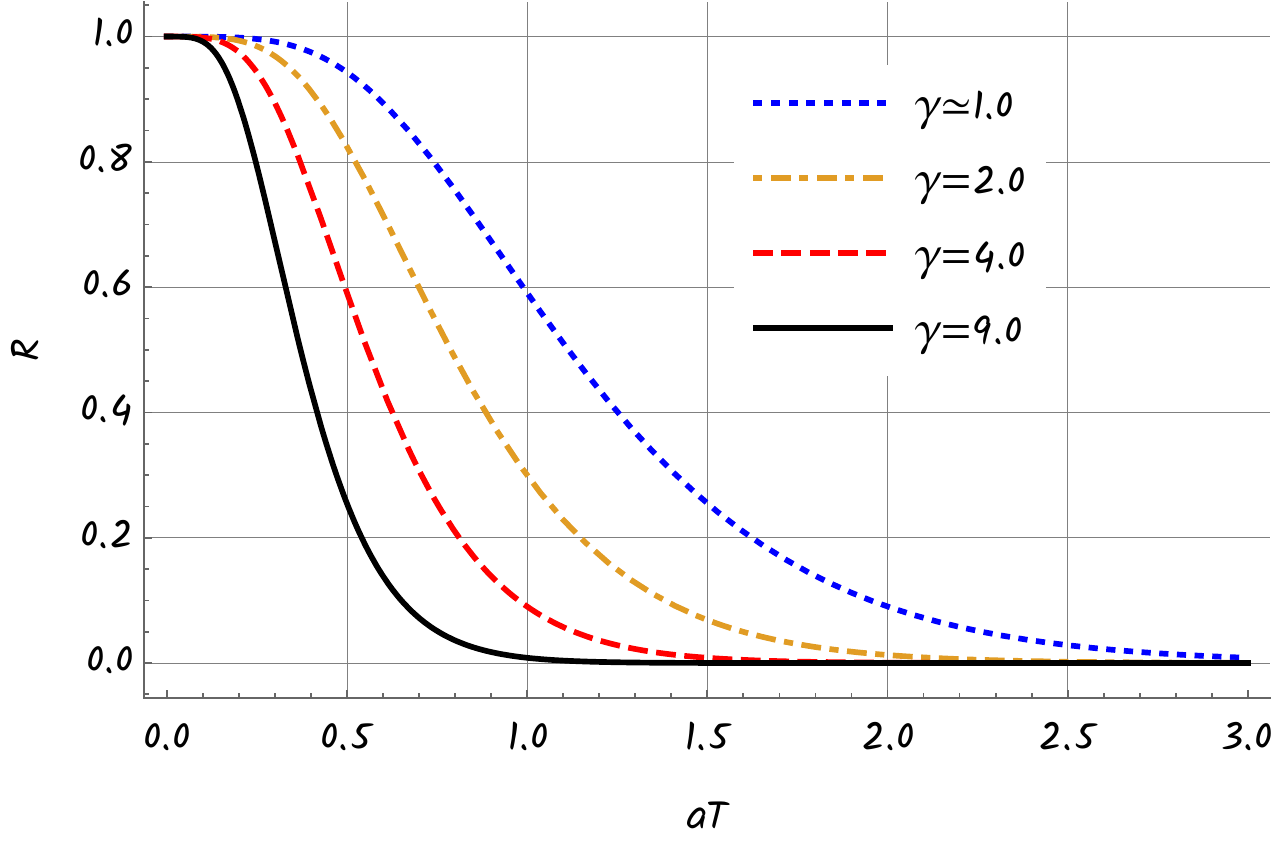}
    \caption{
    Plot of the ratio $R = \mathcal{F}_{\text{ren}}/\mathcal{E}_{\text{cas}}$ in terms of $a\,T$ for several values of the parameter $\gamma$. $R$ decreases with $aT$ and tends to zero when $aT$ goes to infinity. }
    \label{Correction}
\end{figure}

To conduct our analysis, let us rewrite $L$ as ${L = a \sqrt{\gamma}}$ by convenience, where $\gamma = 1 + (b/a)^2$. In Fig. \ref{Correction}, we have plotted the ratio $R$ of the renormalized free energy density, $\mathcal{F}_{\text{ren}}(\beta, L)$, to the Casimir energy density, $\mathcal{E}_{\text{cas}}(L)$, varying with $aT$ for different values of the parameter $\gamma$. 

In each case, the plot shows the ratio $R = \mathcal{F}_{\text{ren}}/\mathcal{E}_{\text{cas}}$ going to $1$ as $T$ approaches zero, as we should expect, and decaying to zero as $T$ approaches infinity. In particular, this decay becomes more pronounced as the parameter $\gamma$ increases, as illustrated by the curve for $\gamma = 9$.   
The curves associated with $\gamma > 2$, which decay to zero faster, correspond to the case where $b$ is greater than $a$, whereas the curve with $\gamma = 2$ illustrates the particular case $b = a$. In the limiting case when $\gamma \simeq 1$ ($b \ll a$), the system exhibits a structure known as a quantum spring, as discussed by \cite{Feng2010} in the context of the scalar Casimir effect. 
 
In what follows, we will analyze the asymptotic limits of temperature corrections to the massless free energy density above. 

\subsubsection{High-temperature limit}

Let us analyze the high-temperature limit, ${\beta \ll L}$, of the final expression \eqref{Massless free energy density}. 
In this case, it is more appropriate to perform the summation in $p$ first, namely
 \begin{align}\label{summation in p}
\frac{1}{\beta^4 } \sum_{p=1}^{\infty} &\frac{\cos(\pi p) }{\left(p^2 + \frac{L^2}{\beta^2} n^2\right)^2 } = -\frac{1}{2\,L^4}\frac{1}{n^4} \nonumber\\
&+ \frac{T^2}{2 L^2}\frac{1}{n^2}\,\text{csch}(n \pi L\, T) \nonumber\\
&\times \left[\coth(n \pi L\, T) + \frac{1}{n \pi L\, T}\right]  . 
\end{align}   
Inserting this summation into Eq. \eqref{Massless free energy density}, we are eventually led to the following expression
\begin{align}\label{Summation in n - high}
  \mathcal{F}_{\text{ren}}\left(\beta, L \right) &=  \frac{4\,T^2}{L^2} \sum_{n=1}^{\infty}\frac{\cos(\pi n)}{n^2}\text{csch}(n \pi L\, T)  \nonumber\\
  &\times \left[\coth(n \pi L\, T) + \frac{1}{n \pi L\, T}\right] .
\end{align}
Note that the first term on the right-hand side in Eq. \eqref{summation in p}, after performing the summation in $n$, gives rise to Casimir energy density associated with massless spinor field \eqref{Massless Casimir energy density} but with the opposite sign. Therefore, the effect of this latter term is entirely compensated by the corresponding one in the free energy density \eqref{Massless free energy density}. Such a natural cancellation between the massless Casimir energy density and its corresponding temperature correction is not unusual. It is an intrinsic characteristic of temperature corrections at the high-temperature limit \cite{Plunien1986}. 

Now, our task is to evaluate the series \eqref{Summation in n - high} at a high-temperature limit, ${\beta \ll L}$ (or equivalently ${L\, T \gg 1}$). Through an asymptotic expansion up to terms of order $\mathcal{O}(e^{-\pi L\, T})$, we arrive at the finite-temperature 
\begin{align}\
\mathcal{F}_{\text{ren}}\left(\beta \ll L \right) \simeq  -\frac{8\,T}{L^3} \left(1 +\pi L\, T\right)\,e^{-\pi L\, T} ,
\end{align}
which is exponentially suppressed at high $T$ and converges to zero as ${T \rightarrow \infty}$, in accordance with Fig. \ref{Correction}. This behavior is expected for a spinor field since, differently from the scalar field \cite{Giulia2021}, it lacks a temperature correction term that is linearly dependent on $T$. We emphasize the need for the free energy density to undergo a finite renormalization by subtracting from it the blackbody radiation contribution, proportional to $T^4$, to obtain the correct classical limit, a free energy density renormalized to zero at very high temperatures. \text{Ref.~\cite{Mostepanenko2011}} found a similar result for temperature corrections associated with the spinor field in the closed Friedmann cosmological model. 

With the renormalized free energy density now available, we can obtain an analytical expression for the renormalized entropy density. Employing the relation \eqref{Casimir entropy}, we have

\begin{align}
    \mathcal{S}_{\text{ren}}(\beta, L)  & =   \frac{4\pi\,T^2}{L^2} \sum_{n=1}^{\infty}\frac{\cos(\pi n)}{n}\text{csch}(n \pi L\, T)  \nonumber\\
  &\times \left[1 - \frac{1}{(n \pi  L\, T)^2} + 2 \,\text{csch}^2(n \pi L\, T) \right. \nonumber\\
 &-  \left. \frac{1}{n \pi L\, T}\coth(n \pi L\, T) \right] . 
\end{align}
The corresponding asymptotic expansion in the high-temperature regime, ${L\, T \gg 1}$, decays exponentially
with the temperature $T$
\begin{align}
    \mathcal{S}_{\text{ren}}\left(\beta \ll L \right)  \simeq   -\frac{8}{\pi L^3 } \left[1 + \pi L\,T  - (\pi L\, T)^2 \right]\,e^{-\pi L\, T} .
\end{align}
Note that the lack of a classical term proportional to $T$ in the free energy density results in the Casimir entropy density approaching zero at very high temperatures, which differs from the scalar case where it is dominated by a constant term \cite{Giulia2021}.

\subsubsection{Low-temperature limit}

Let us now consider the asymptotic expansion of the expression \eqref{Massless free energy density} in the low-temperature regime, where ${\beta \gg L}$, or equivalently ${L\, T \ll 1}$. To accomplish this, as previously mentioned, we shall perform the summation in $n$ first, providing
\begin{align}\label{summation in n}
\frac{1}{\beta^4 } \sum_{n=1}^{\infty} &\frac{\cos(\pi p) }{\left(p^2 + \frac{L^2}{\beta^2} n^2\right)^2 } = -\frac{T^4}{2\,\pi^2}\frac{1}{p^4} \nonumber\\
&+ \frac{4\,T^2}{L^2}\frac{1}{p^2} \,\text{csch}\left(\frac{p \pi}{L\, T}\right) \nonumber\\
&\times \left[ \frac{L\, T}{p \pi} +\coth\left(\frac{p \pi}{L\, T}\right)\right] . 
\end{align} 
Substituting it back in Eq. \eqref{Massless free energy density}, we get
\begin{align}\label{summation in p - Low}
 \mathcal{F}_{\text{ren}}\left(\beta, L \right) &\simeq  -\frac{7}{2}\,\frac{\pi^2}{90\,L^4} +\frac{7}{2} \frac{\pi^2}{90} \, T^4 \nonumber\\
&+ \frac{4\,T^2}{L^2} \sum_{p=1}^{\infty} \frac{\cos(\pi p)}{p^2} \,\text{csch}\left(\frac{p \pi}{L\, T}\right) \nonumber\\
&\times \left[ \frac{L\, T}{p \pi} +\coth\left(\frac{p \pi}{L\, T}\right)\right] , 
\end{align}  
which at the low-temperature regime up to terms of the order $\mathcal{O}(e^{-\frac{\pi}{L\, T}})$, presents the following free energy density asymptotic behavior
\begin{align}\label{summation in p - Low}
 \mathcal{F}_{\text{ren}}\left(\beta \gg L \right) &\simeq   -\frac{7}{2}\,\frac{\pi^2}{90\,L^4} + \frac{7}{2} \frac{\pi^2}{90} \, T^4 \nonumber\\
 & - \frac{8\, T^2}{L^2} \left(1 + \frac{L\, T}{\pi}\right)\,e^{-\frac{\pi}{L\, T}} . 
\end{align}  
Let us make a few remarks comparing our findings and those reported in \cite{Giulia2021} for the low-temperature behavior of the free energy associated with the scalar field under helix topology. Apart from the additional $T^3$ term for temperature correction observed in the scalar case, they differ by constant multiplicative factors that naturally arise because of the spinor degrees of freedom. Note that at small $T$, the above asymptotic expansion is dominated by the first term, the massless Casimir energy density at zero temperature, Eq. \eqref{Massless Casimir energy density}, as expected \cite{Mostepanenko2011, Herondy2023, Basil1978}.

The entropy density can be obtained by inserting Eq. \eqref{summation in p - Low} into Eq. \eqref{Casimir entropy}, providing
\begin{align}
    \mathcal{S}_{\text{ren}}(\beta, L)  & =  -\frac{7\pi^2}{45}\, T^3 \nonumber\\
    & -\frac{4\pi}{L^3} \sum_{p=1}^{\infty}\frac{\cos(\pi p)}{p}\,\text{csch}\left(\frac{p \pi}{L\, T}\right)  \nonumber\\
  &\times \left[1 + 3\left(\frac{L\, T}{p \pi}\right)^2 + 2\,\text{csch}^2\left(\frac{p \pi}{L\, T}\right) \right. \nonumber\\
 &-  \left. \frac{3 L\, T}{p \pi}\coth\left(\frac{p \pi}{L\, T}\right) \right] . 
\end{align}
Its corresponding asymptotic expansion in the low-temperature limit, where ${L\, T \ll 1}$, is found to be
\begin{align}
    \mathcal{S}_{\text{ren}}
    \left(\beta \gg L\right) &\simeq  \frac{7\pi^2}{45}\, T^3 \nonumber\\
    &+\frac{8\pi}{L^3} \left[1 + \frac{3 L\, T}{\pi}  + 3\left(\frac{L\, T}{\pi}\right)^2 \right]\,e^{-\frac{\pi}{L\, T}} .
\end{align}
As expected, the above expression tends to zero as the temperature approaches zero. It implies that the entropy density for a massless spinor field satisfying an anti-periodic condition along the compact vector satisfies the third law of thermodynamics (the Nernst heat theorem) \cite{Landau1980}. 
 

\section{Conclusion}\label{Conclusion}

In the present work, we have investigated the thermal Casimir effect associated with a massive spinor field defined on a four-dimensional flat space with a circularly compactified dimension. The periodicity represented by $\mathbb{S}^1$ is oriented not along a coordinate axis as usual, but along a vector $\boldsymbol{L}$ belonging to the $xy$-plane, Eq. \eqref{compact vector}. This geometry introduces a topological constraint inducing a spatial anti-periodic boundary condition on the spinor field, Eq. \eqref{Helix condition}, which modifies the vacuum fluctuations, producing the Casimir effect. Imposing this boundary condition led to the discrete eigenvalues for the momentum along vector $\boldsymbol{L}$, Eq. \eqref{kx ky relation}, allowing for determining explicitly the eigenvalues \eqref{spatial eigenvalue}. They are used to construct the generalized zeta function for the spinor field and thus remove the formal divergences involved in the Casimir effect. 

These divergencies were introduced by the Dirac operator determinant in the partition function originating from the infinite product over eigenvalues, Eq. \eqref{Functional determinant}.
This divergence was encoded into the generalized zeta function employing the important relation connecting it with the partition function, Eq.  \eqref{Log Z}. It was analyzed from the asymptotic behavior of the spinor heat kernel function, Eq. \eqref{small t expansion}, and removed in the renormalization scheme by subtraction of the divergent contribution associated with non-null heat kernel coefficients. A rather peculiar aspect of the zeta function regularization prescription is related to the existence of ambiguities. Such ambiguities appear whenever the mass-dependent $c_2(m)$ heat kernel coefficient is nonvanishing, Eq. \eqref{c_2(m) heat kernel coefficient}, due to natural dependence on parameter $\mu$, Eq. \eqref{Effetive energy}. 
For the geometry presented here, $c_0$ was the only non-null heat kernel coefficient, Eq. \eqref{c0 heat kernel coefficient}, associated with the Euclidean heat kernel contribution, Eqs. \eqref{Euclidean heat kernel function} and \eqref{Euclidean vacuum-free energy}. In order to derive physically meaningful expressions, all contributions associated with $c_0$ were dropped to ensure that the renormalization procedure is unique and thus obtain an unambiguous spinor vacuum free energy.
Besides that, since $c_0$ is multiplied by mass with a positive exponent, we adopt an additional requirement that vacuum energy should be renormalized to zero for large masses.

We outline all the mathematical machinery required for computing the vacuum-free energy density, starting with the construction of the partition function for the spinor field through Euclidean path integrals. In this Euclidean approach, we find closed and analytical expressions for the vacuum free energy density associated with the spinor field in thermal equilibrium at finite temperature $T = \beta^{-1}$, satisfying anti-periodic conditions in the imaginary time $t$ and along vector $\boldsymbol{L}$. This energy density can be expressed as a summation of the zero-temperature Casimir energy density, Eq. \eqref{Massive Casimir}, and temperature correction terms, Eq. \eqref{Renormalized vacuum-free energy}, which generalize the results presented in Refs. \cite{Bellucci2009, Xiang2011}. We also analyzed the high- and low-temperature asymptotic limits, which agree entirely with the curves shown in Fig. \ref{Correction}. The ratio of the renormalized free energy density to the Casimir energy density goes to $1$ as $T$ approaches zero and decays to zero as $T$ approaches infinity. At high temperatures, in particular, we have shown that the $c_0$ coefficient gives rise to the Stefan-Boltzmann law, proportional to $T^4$. Although not divergent, this quantum term was subtracted in the renormalization procedure to obtain a correct classical contribution in this limit. Also, the free energy density does not possess a classical limit at high temperatures. Except for this classical limit, all our results for spinor fields differ from the ones for scalar fields by constant multiplicative factors that naturally arise because of the spinor degrees of freedom. 
Finally, our analysis confirms that the entropy density agrees with the Nernst heat theorem.

\section{Acknowledgments}

J. V. would like to thank Funda\c{c}\~{a}o de Amparo a Ci\^{e}ncia e Tecnologia do Estado de Pernambuco (FACEPE), for their partial financial support. A. M. acknowledges financial support from the Brazilian agencies Conselho Nacional de Desenvolvimento Científico e Tecnológico (CNPq), grant no. 309368/2020-0 and Coordenação de Aperfeiçoamento de Pessoal de Nível Superior (CAPES). H. M. is partially supported by CNPq under grant no. 308049/2023-3. L. F. acknowledges support from CAPES, financial code 001. 





\begin{thebibliography}{9}

\bibitem{Casimir}H. B. G. Casimir, \textit{On the Attraction Between Two Perfectly Conducting Plates, Indag. Math.} {\bf10}, 261  (1948).

\bibitem{Sparnaay} M. J. Sparnaay, Measurements of attractive forces between flat plates, Physica {\bf24}, 751 (1958).

\bibitem{Bressi2002} G. Bressi, G. Carugno, R. Onofrio, and G. Ruoso, \textit{Measurement of the Casimir force between parallel metallic surfaces}, Phys. Rev. Lett., \textbf{88}, 041804 (2002).

\bibitem{Lamoreaux}  S. K. Lamoreaux, Erratum: \textit{Demonstration of the casimir force in the 0.6 to 6 µm range},  Phys. Rev. Lett. \textbf{81}, 5475 (1998).

\bibitem{Lamoreauxx} S. K. Lamoreaux, \textit{Demonstration of the Casimir force in the 0.6 to 6 micrometers range}, Phys. Rev. Lett. {\bf78}, 5 (1997).
Erratum: Phys.Rev.Lett. {\bf 81}, 5475 (1998).

\bibitem{Mohideen}U. Mohideen and A. Roy, \textit{Precision measurement of the Casimir force from 0.1 to 0.9 micrometers}, Phys. Rev. Lett. {\bf81}
 4549 (1998).


\bibitem{Bordag} M. Bordag, U. Mohideen, and V. M. Mostepanenko, \textit{New developments in the Casimir effect, Phys. Rept.} {\bf353}, 1 (2001).

\bibitem{Mostepanenko}V. M. Mostepanenko and N. N. Trunov, \textit{The Casimir effect and its applications} (Clarendon Press, Oxford, New York,
1997). 

\bibitem{Ford1975} L. H. Ford, \textit{Quantum vacuum energy in general relativity}, Phys. Rev. D \textbf{11}, 3370 (1975).

\bibitem{Dowker 1976} J. S. Dowker and R. Critchley, \textit{Covariant Casimir calculations}, J. Phys. A \textbf{9}, 535 (1976).

\bibitem{Dowker1978} J. S. Dowker, \textit{Thermal properties of Green's functions in Rindler, de Sitter, and Schwarzschild spaces}, Phys. Rev. D \textbf{18}, 1856 (1978).

\bibitem{Appelquist1983} T. Appelquist and A. Chodos, \textit{Quantum Effects in Kaluza-Klein Theories}, Phys. Rev. Lett. \textbf{50}, 141 (1983).

\bibitem{Hosotani1983} Y. Hosotani, \textit{Dynamical gauge symmetry breaking as the Casimir effect}, Phys. Lett. B \textbf{129}, 193 (1983).

\bibitem{Brevik2002} I. Brevik, K.A. Milton, S.D. Odintsov, \textit{Entropy Bounds in $R\times S^3$ Geometries}, Ann. Phys. \textbf{302}, 120 (2002). 

\bibitem{Zhang2015} A.  Zhang, \textit{Thermal Casimir Effect in Kerr Space-time}, Nuclear Physcis B \textbf{898}, 2020, (2015).

\bibitem{Henke2018}C.  Henke, \textit{Quantum vacuum energy in general relativity}, Eur. Phys. J. C \textbf{78}, 126 (2018).

\bibitem{Bradonjic2009} K. Bradonjic, J. Swain, A. Widom, and Y. Srivastava, \textit{The casimir effect in biology: The role of molecular quantum
electrodynamics in linear aggregations of red blood cells}, Journal of Physics: Conference Series {\bf161}, 012035 (2009).

\bibitem{Peng2018} Peng Liu and Ji-Huan He, \textit{Geometric potential: An explanation of nanofibers wettability}, Thermal Science {\bf22}, 33 (2018). 

\bibitem{Pawlowski2013} P. Pawlowski and P. Zielenkiewicz, \textit{The quantum casimir effect may be a universal force organizing the bilayer structure of the cell membrane}, The Journal of membrane biology {\bf246}, 383 (2013).

\bibitem{Gambassi2009} A. Gambassi, \textit{The Casimir effect: From quantum to critical fluctuations}, J. Phys. Conf. Ser. {\bf161}, 012037 (2009).

\bibitem{Machta2012} B. B. Machta, S. L. Veatch, and J. P. Sethna, \textit{Critical casimir forces in cellular membranes}, Phys. Rev. Lett. {\bf109}, 138101 (2012).


\bibitem{Klimchitskaya} G. L. Klimchitskaya, U. Mohideen and V. M. Mostepanenko, \textit{The Casimir force between real materials: Experiment and theory}, Rev. Mod. Phys. \textbf{81}, 1827 (2009).

\bibitem{Milton2019} K. Milton and I. Brevik, \textit{Casimir Physics Applications}, Symmetry \textbf{11}, 201 (2019).

\bibitem{Milton2001} K. A. Milton, \textit{The Casimir Effect: Physical Manifestation of Zero-Point Energy} (World Scientific, Singapore, 2001).

\bibitem{Klimchitskaya2009} M. Bordag, G. L. Klimchitskaya, U. Mohideen, and V. M. Mostepanenko, \textit{Advances in the Casimir effect} (Oxford University, Press, Oxford, 2009), Vol. 145.


\bibitem{Stokes2015} A. Stokes and R. Bennett, \textit{The Casimir effect for fields with arbitrary spin}, Annals Phys. \textbf{360}, 246 (2015).


\bibitem{Farina2006} C. Farina, \textit{The Casimir Effect: Some Aspects}. Brazilian Journal of Physics \textbf{36}, 1137 (2006).

\bibitem{Muniz2018} C. R. Muniz, M. O. Tahim, M. S. Cunha and H. S. Vieira, \textit{On the global Casimir effect in the Schwarzschild spacetime}, JCAP \textbf{1801}, 006 (2018).

\bibitem{Cunha2016} M. S. Cunha, C. R. Muniz, H. R. Christiansen, V. B. Bezerra, \textit{Relativistic Landau levels in the rotating cosmic string spacetime}, Eur. Phys. J. C \textbf{76}, 512 (2016).

\bibitem{Mobassem2014} S. Mobassem, \textit{Casimir effect for massive scalar field}, Mod. Phys. Lett. A \textbf{29}, 1450160 (2014).

\bibitem{Pereira2017} S. H. Pereira, J. M. Hoff da Silva and R. dos Santos, \textit{Casimir effect for Elko fields}, Mod. Phys. Lett. A \textbf{32}, 1730016 (2017).

\bibitem{Bytsenko2005} A. A. Bytsenko, M. E. X. Guimarães and V. S. Mendes, \textit{Casimir Effect for Gauge Fields in Spaces with Negative Constant Curvature}, Eur. Phys. J. C \textbf{39}, 249 (2005). 

\bibitem{Chernodub2018} M. N. Chernodub, V. A. Goy, A. V. Molochkov and Ha Huu Nguyen, \textit{Casimir effect in Yang-Mills theory}, Phys. Rev. Lett. \textbf{121}, 191601 (2018)

\bibitem{Edery2007} A. Edery, \textit{Casimir piston for massless scalar fields in three dimensions, Phys. Rev.} D {\bf75} (2007) 105012, [hep-th/0610173].

\bibitem{Photon2001} E. Ponton and E. Poppitz, \textit{Casimir Energy and Radius Stabilization in Five and Six Dimensional Orbifolds}, JHEP \textbf{0106}, 019 (2001).

\bibitem{Cartan1966} E. Cartan, \textit{The Theory of Spinors}, Dover (1966).

\bibitem{Benn1987} I. Benn and R. Tucker, \textit{An Introduction to spinors and geometry with applications in physics}, Adam
Hilger, Bristol (1987).

\bibitem{JoasBook2019} J. Venâncio, \textit{The spinorial formalism}. Lambert Academic Publishing, Germany (2019).

\bibitem{Oikonomou2010} V. K. Oikonomou and N. D. Tracas, \textit{Slab Bag Fermionic Casimir effect, Chiral Boundaries and Vector Boson-Majorana Fermion Pistons}, Int. J. Mod. Phys. A \textbf{25}, 5935 (2010).

\bibitem{Cheng2010} H. Cheng, \textit{Casimir effect for parallel plates involving massless Majorana fermions at finite temperature}, Phys. Rev. D \textbf{82}, 045005 (2010).

\bibitem{Cavalcanti2004} R. M. Cavalcanti, \textit{Casimir force on a piston}, Phys. Rev. D \textbf{69}, 065015 (2004).

\bibitem{Elizalde2008} E. Elizalde, \textit{Zeta function methods and quantum fluctuations}, J. Phys. A \textbf{41}, 304040 (2008).

\bibitem{Elizalde1995} E. Elizalde, \textit{Ten Physical Applications of Spectral Zeta Functions} (Springer, New York, 1995).

\bibitem{Hawking1977} S. W. Hawking, \textit{Zeta function regularization of path integrals in curved spacetime}, Commun. math. Phys. \textbf{55}, 133 (1977).

\bibitem{Elizalde2012} E. Elizalde, \textit{Zeta function regularization in Casimir effect calculations and J.S. Dowker's contribution},
Int. J. Mod. Phys. A \textbf{27}, 1260005 (2012). 

\bibitem{Elizalde1994} E. Elizalde, S. D. Odintsov, A. Romeo, A. A. Bytsenko, and S. Zerbini, \textit{Zeta regularization techniques with applications}, World Scientific (1994). 

\bibitem{Basil1978} M. Basil Altaie and J. S. Dowker, \textit{Spinor fields in an Einstein universe: Finite-temperature effects}, Phys. Rev. D \textbf{18}, 3557 (1978).  

\bibitem{Plunien1986} G. Plunien, B. Muller and W. Greiner, \textit{The Casimir effect}, Phys. Rept. \textbf{134}, 87 (1986).

\bibitem{Kulikov1988} I. K. Kulikov, P. I. Pronin. \textit{Finite temperature contributions to the renormalized energy-momentum tensor for an arbitrary curved space-time}. Czech J Phys \textbf{38}, 121 (1988).

\bibitem{Maluf2020} R. V. Maluf, D. M. Dantas and C. A. S. Almeida, \textit{The Casimir effect for the scalar and Elko fields in a Lifshitz-like field theory},  Eur. Phys. J. C \textbf{80}, 442 (2020).

\bibitem{Kac1966} M. Kac, \textit{Can one hear the shape of a drum?} Amer. Math. Monthly \textbf{73}, 1 (1966).

\bibitem{Bordag2000} M. Bordag, \textit{Ground state energy for massive fields and renormalization}, Commun. Mod. Phys., Part D \textbf{1}, 347 (2000).

\bibitem{Vassilevich2003} D. V. Vassilevich, \textit{Heat kernel expansion: user's manual}, Phys. Rept. \textbf{388}, 279 (2003). 

\bibitem{Kirstein2010} K. Kirstein, \textit{Basic zeta functions and some applications in physics}, 2010. arXiv:1005.2389


\bibitem{Branson1992A} T. Branson and P. Gilkey, \textit{Residues of the eta function for an operator of
Dirac type}, J. Funct. Anal. \textbf{108}, 47 (1992).

\bibitem{Branson1992B} T. Branson and P. Gilkey, \textit{Residues of the eta function for an operator of
Dirac type with local boundary condtitons}, Differential Geom. Appl. \textbf{2}, 249 (1992).

\bibitem{Chodos1974} A. Chodos, R. L. Jaffe, K. Johnson, C. B. Thorn, and V. F. Weisskopf, \textit{New extended model of hadrons}, Phys. Rev. D \textbf{9}, 3471 (1974).

\bibitem{Johnson1975} K. Johnson, \textit{The M.I.T. bag model}, Acta Phys. Polon. B \textbf{6}, 865 (1975).

\bibitem{Arrizabalaga2017} N. Arrizabalaga, L. Le Treust and N. Raymond, \textit{On the MIT bag model in the non-relativistic limit}, Commun. Math. Phys. \textbf{354}, 641 (2017). 

\bibitem{Mamayev1980} S. G. Mamayev and N. N. Trunov, \textit{Vacuum expectation values of the energy-momentum tensor of quantized fields on manifolds with different topologies and geometries. III}, Sov. Phys. J. \textbf{23}, 551, (1980).

\bibitem{Erdas2011} A. Erdas, \textit{Finite temperature Casimir effect for massless Majorana fermions in a magnetic field}, Phys. Rev. D \textbf{83}, 025005 (2011). 

\bibitem{Elizalde2012Maj} E. Elizalde et al., \textit{The Casimir energy of a massive fermionic field confined in a $(d+1)$-dimensional slab-bag}, International Journal of Modern Physics A \textbf{18}, 1761 (2012).

\bibitem{Elizalde1998} E. Elizalde, M. Bordag and K. Kirsten, \textit{Casimir energy for a massive fermionic quantum field with a spherical boundary}, J. Phys. A: Math. Gen. \textbf{31}, 1743 ( 1998). 


\bibitem{Bender1976} C. M Bender and P. Hays, \textit{Zero-point energy of fields in a finite volume}, Phys. Rev. D \textbf{14}, 2622 (1976).

\bibitem{Fucci2023} G. Fucci and C. Romaniega, \textit{Casimir energy for spinor fields with $\delta$-shell potentials}, J. Phys. A: Math. Theor. \textbf{56}, 265201 (2023). 

\bibitem{Pereira2019} S. H. Pereira and R. S. Costa, \textit{Partition function for a mass dimension one fermionic field and the dark matter halo of galaxies}, Mod. Phys. Lett. A \textbf{34}, 1950126 (2019).

\bibitem{Xin2011} Xiang-hua Zhai, Xin-zhou Li, Chao-Jun Feng, \textit{Casimir effect with a helix torus boundary condition}, Mod. Phys. Lett. A \textbf{26}, 1953 (2011).

\bibitem{Farias2020} K. E. L. de Farias and H. F. Santana Mota, \textit{Quantum vacuum fluctuation effects in a quasi-periodically identified conical
spacetime, Phys. Lett.} B {\bf807} (2020) 135612, [arXiv:2005.03815].






\bibitem{Xin-zhou} Chao-Jun Feng and Xin-zhou Li, \textit{Quantum Spring from the Casimir Effect, Phys. Lett.} B {\bf691} (2010) 167–172, [arXiv:1007.2026].


\bibitem{Zhai} Xiang-hua Zhai, Xin-zhou Li, Chao-Jun Feng, \textit{The Casimir force of Quantum Spring in the (D+1)-dimensional spacetime,
Mod. Phys. Lett.} A {\bf26} (2011) 669–679, [arXiv:1008.3020].

\bibitem{Li} Chao-Jun Feng and Xin-zhou Li, \textit{Quantum Spring, Int. J. Mod. Phys. Conf. Ser.} {\bf7} (2012) 165–173, [arXiv:1205.4475].

\bibitem{HerondyJunior2015} H. F. Mota and V. B. Bezerra, \textit{Topological thermal Casimir effect for spinor and electromagnetic fields}, Phys. Rev. D \textbf{92}, 124039 (2015).

\bibitem{Mohammadi2022} K. E. L. de Farias, A. Mohammadi, and H. F. Santana Mota, \textit{Thermal Casimir effect in a classical liquid in a quasi-periodically identified conical spacetime}, Phys. Rev. D \textbf{105}, 085024 (2022).

\bibitem{Herondy2023} H. Mota, \textit{Vacuum energy, temperature corrections and heat kernel coefficients in $(D+1)$-dimensional spacetimes with nontrivial topology}, 2023. arXiv:2312.01909


\bibitem{Mostepanenko2011} V. B. Bezerra, V. M. Mostepanenko, H. F. Mota, and C. Romero, \textit{Thermal Casimir effect for neutrino and electromagnetic fields in the closed Friedmann cosmological model}, Phys. Rev. D \textbf{84}, 104025 (2011).

\bibitem{Birrell1980} N. D. Birrell and L. H. Ford, \textit{Renormalization of Self-Interacting Scalar Field Theories in a Nonsimply Connected Spacetime}. Physical Review D \textbf{22}, 330 (1980). 


\bibitem{Xiang2011} Xiang-hua Zhai, Xin-zhou Li and Chao-Jun Feng, \textit{Fermionic Casimir effect with helix boundary condition}, Eur. Phys. J. C \textbf{71}, 1654 (2011).

\bibitem{Bellucci2009} S. Bellucci, A.A. Saharian, \textit{Fermionic Casimir effect for parallel plates in the presence of compact dimensions
with applications to nanotubes}, Phys. Rev. D \textbf{80}, 105003 (2009).

\bibitem{Giulia2021} G. Aleixo, H. F. Santana Mota, \textit{Thermal Casimir effect for the scalar field in flat spacetime under a helix boundary condition}, Phys. Rev. D \textbf{104}, 045012 (2021).

\bibitem{Chao-Jun Feng} Chao-Jun Feng, Xin-Zhou Li, Xiang-Hua Zhai, \textit{Casimir Effect under Quasi-Periodic Boundary Condition Inspired by Nanotubes}, Mod. Phys. Lett A {\bf29}, 1450004 (2014

\bibitem{Liu2022} B. Liu, C. L. Yuan, H. L. Hu, et al. \textit{Dynamically actuated soft heliconical architecture via frequency of electric fields}. Nat Commun \textbf{13}, 2712 (2022).

\bibitem{Greenfeld2020} I. Greenfeld, I. Kellersztein and H. D. Wagner, \textit{Nested helicoids in biological microstructures}. Nat Commun \textbf{11}, 224 (2020).

\bibitem{Elizalde1990} E. Elizalde and A. Romeo, \textit{Heat-kernel approach to zeta function regularization of the Casimir effect for domains with curved boundaries}, Int. J. Mod. Phys. A \textbf{5}, 1653, (1990).

\bibitem{Seeley1967} R.T. Seeley, \textit{Complex powers of an elliptic operator}, Amer. Math. Soc. Proc. Symp. Pure Math. \textbf{10}, 288 (1967).

\bibitem{Vassilevich2011} D. V. Vassilevich, \textit{Operators, Geometry and Quanta}, Springer (2011). 

\bibitem{Gilkey1995} P. Gilkey, \textit{Invariance Theory, the Heat Equation, and the Atiya-Singer Index Theorem}. CRC Press, Boca Raton, FL, 1995.

\bibitem{Kulikov1989} I. K. Kulikov and P. I. Pronin, \textit{Topology and chiral symmetry breaking in four-fermion interaction}, Acta Phys. Polon. B \textbf{20}, 713 (1989).  

\bibitem{Ahmadi2005} N. Ahmadi and M. Nouri-Zonoz, \textit{Massive spinor fields in flat spacetimes with non-trivial topology}, 	Phys. Rev. D \textbf{71}, 104012 (2005).

\bibitem{Joas2017} J. Ven\^ancio and C. Batista, \textit{Separability of the Dirac equation on backgrounds that are the direct product of bidimensional spaces}, Physical Review D \textbf{95}, 084022 (2017).

\bibitem{Mohammadi2013} S. S. Gousheh, A. Mohammadi, and L. Shahkarami, \textit{Casimir energy for a coupled fermion-kink system and its stability}, Phys. Rev. D \textbf{87}, 045017 (2013).

\bibitem{Mohammadi2015} A. Mohammadi, E. R. Bezerra de Mello and A. A. Saharian, \textit{Induced fermionic currents in de Sitter spacetime in the presence of a compactified cosmic string}, Class. Quantum Grav. \textbf{32}, 135002 (2015).

\bibitem{Mohammadi2020} E. A. F. Bragança, E. R. Bezerra de Mello, and A. Mohammadi, \textit{Induced fermionic vacuum polarization in a de Sitter spacetime with a compactified cosmic string}, Phys. Rev. D \textbf{101}, 045019 (2020).












\bibitem{Bytsenko1992} A. Bytsenko, L. Vanzo, S. Zerbini, \textit{Zeta-function regularization approach to finite temperature effects in Kaluza-Klein space-times}, Mod. Phys. Lett. A \textbf{7}, 2669 (1992). 

\bibitem{Zerbini1993} S. Zerbini, \textit{Spinor fields on $2 + 1$ topologically nontrivial spacetime}, Letters in Mathematical Physics \textbf{27}, 19 (1993).

\bibitem{Landau1980} L. D. Landau and E. M. Lifshitz, \textit{Statistical Physics, Part I}. Pergamon Press, Oxford, 1980.


\bibitem{Feng2010}  C.J. Feng, X.Z. Li, \textit{Quantum spring from the Casimir effect}, Phys. Lett. B \textbf{691}, 167 (2010).

\bibitem{Abramowitz1972} M. Abramowitz and I. A. Stegun, \textit{Handbook of Mathematical Functions with Formulas, Graphs, and Mathematical Tables} (Dover, New York, 1972).















\end{thebibliography}
\end{document}